\documentclass[a4paper,11pt]{article}
\usepackage{dcolumn}
\usepackage{psfrag}
\usepackage{wrapfig}
\usepackage{subfigure}
\usepackage{makeidx}
\usepackage{epsf}
\usepackage{color}
\usepackage{amsmath, amssymb, amsfonts} % 必须在 cases 之前，否则 cases 会抢先定义 \subequations
\usepackage{makecell} %表格内换行
\usepackage{float}

\usepackage[left=2.4cm,right=2.4cm,top=2.5cm,bottom=2.5cm]{geometry}
\usepackage{cases}
\usepackage{multirow}
\usepackage{mathrsfs}
\usepackage{booktabs,array}
\usepackage{placeins}
\usepackage{graphicx}
\usepackage{hyperref}
\usepackage{bm}
\usepackage{xcolor}
\usepackage{ulem} % 提供 \sout{} 命令用于删除线

\hypersetup{
  colorlinks=true,
  linkcolor=blue!55!black,
  citecolor=blue!55!black,
  urlcolor=blue!55!black,
  pdfauthor={},
  pdftitle={Exceptional Points in Scalar Quasinormal Modes of Rastall String-Fluid Black Holes}
}

\AtBeginDocument{%
  \setlength{\oddsidemargin}{2.4cm}%
  \addtolength{\oddsidemargin}{-1in}%
  \setlength{\evensidemargin}{\oddsidemargin}%
  \setlength{\textwidth}{\paperwidth}%
  \addtolength{\textwidth}{-4.8cm}%
  \setlength{\columnwidth}{\textwidth}%
  \setlength{\linewidth}{\textwidth}%
  \setlength{\hsize}{\textwidth}%
  \setlength{\topmargin}{2.5cm}%
  \addtolength{\topmargin}{-1in}%
  \setlength{\textheight}{\paperheight}%
  \addtolength{\textheight}{-5cm}%
}


\newcommand{\ii}{\mathrm{i}}
\newcommand{\eR}{\varepsilon_{\mathrm R}}
\newcommand{\eb}{\epsilon_{\mathrm b}}
\newcommand{\ebc}{\epsilon_{\mathrm b,c}}
\newcommand{\scri}{\mathscr{I}}
\newcommand{\dd}{\mathrm{d}}
\DeclareMathOperator{\sech}{sech}

\begin{document}

\title{Exceptional Points in Scalar Quasinormal Modes of Rastall String-Fluid Black Holes}

\author{Ming Zhang$^{1}$\footnote{ zhangming@xaau.edu.cn}, 
Xiao-Mei Wang$^{1}$\footnote{ wangxiaomei@xaau.edu.cn},
Ya-Nan Wang$^{1}$\footnote{ wangyanan@xaau.edu.cn}, 
Miao Wang$^{1}$\footnote{ wangmiao@xaau.edu.cn}, \\
Lei Zhang$^{1}$\footnote{ zhanglei@xaau.edu.cn}, 
Xufen Zhang$^{2}$\footnote{ xfzhangyzu@126.com}, 
and De-Cheng Zou$^{3}$\footnote{Corresponding author: dczou@jxnu.edu.cn}\\[1ex]
\parbox{\dimexpr\textwidth-2\tabcolsep\relax}{\centering\normalsize\itshape
$^{1}$Faculty of Science, Xihang University, Xi'an 710077 China\\
$^{2}$Key Laboratory of Low Dimensional Quantum Structures and Quantum Control of Ministry of Education,
Synergetic Innovation Center for Quantum Effects and Applications, and Department of Physics,
Hunan Normal University, Changsha, Hunan 410081, China \\
$^{3}$School of Physics, Jiangxi Normal University, Nanchang 330022, China}}

\date{\today}

\maketitle

\begin{abstract}
In this paper, we study exceptional points (EPs) in scalar quasinormal modes(QNMs) of Rastall string-fluid black holes by adding a localized P\"oschl--Teller perturbation to the scalar potential, with the bump position $a$ and amplitude $\eb$ serving as independent controls. We solve the hyperboloidal problem with two-domain Chebyshev collocation and arbitrary-precision linear algebra, and use the complex discriminant, eigenvector coalescence, Jordan-chain residuals, square-root splitting and closed-loop mode exchange as complementary tests of defectiveness. For the regular $A=1$, $\eR=-1$ branch, the two-parameter search locates a resolution-stable second-order EP from $N=24$ to $N=60$, while constrained semidiscrete evolution gives the Jordan response $(C+Dt)e^{\ii\omega_{\rm EP}t}$. For the logarithmic $A=2$ branch, defective points occur in each finite truncation but their coordinates do not converge, so we make no continuum claim for this branch.
\end{abstract}


\section{Introduction}
\label{sec:introduction}

Exceptional points (EPs) are defective degeneracies of non-Hermitian spectral problems. At a second-order EP, two eigenvalues and their right eigenvectors coalesce. The algebraic multiplicity is two, but only one eigendirection remains \cite{Kato1995,Heiss:1999alz,Heiss:2012dx}, so the local operator contains a nontrivial $2\times2$ Jordan block. The local geometry of an EP determines the response to a perturbation. Near a second-order EP, the two eigenvalues generally split as
\begin{equation}
 \omega_{\pm}(\bm p)=\omega_{\rm EP}
 \pm c\sqrt{\bm v\cdot(\bm p-\bm p_{\rm EP})}+\cdots,
 \label{eq:sqrt}
\end{equation}
Here $\bm p$ denotes the control-parameter vector, and equality of two complex eigenvalues imposes two real conditions, so an isolated second-order EP generally requires two independent real controls. Encircling the branch point exchanges the two modes. Eigenvector coalescence also produces self-orthogonality and enhanced spectral sensitivity \cite{El-Ganainy:2018ksn,Bergholtz:2019deh,Ashida:2020dkc}. A small frequency gap alone is not sufficient evidence for an EP because an avoided crossing can produce the same observation.

This structure is relevant to black-hole quasinormal modes(QNMs), which describe the ringdown spectrum \cite{Berti:2025hly}. QNMs satisfy ingoing conditions at the horizon and outgoing conditions at infinity, so their complex frequencies belong to a dissipative, non-Hermitian problem \cite{Kokkotas:1999bd,Nollert:1999ji,Berti:2009kk,Konoplya:2011qq}. At an EP, the Green function has a second-order pole and the time-domain response contains a Jordan term proportional to $te^{+\ii\omega_{\rm EP}t}$ \cite{Yang:2025dbn,PanossoMacedo:2025xnf}. Recent black-hole calculations have reported EPs or closely related resonances in Kerr and Kerr--Newman spacetimes, as well as in hairy and perturbed black-hole models \cite{Motohashi:2024fwt,Cavalcante:2024swt,Cavalcante:2024kmy,Cavalcante:2025abr,Chen:2025sbz,Nakamoto:2026lyo,Cavalcante:2026exl,Cheng:2026eph,Yang:2025dbn,Cao:2025afs,Wu:2025wbp}. Ringdown studies also show that the mean frequency of a resonant pair can remain well conditioned near an EP \cite{PanossoMacedo:2025xnf,Kubota:2025hjk,Oshita:2025ibu,Lo:2025njp,Takahashi:2025uwo}, while pseudospectral analyses establish the sensitivity of black-hole QNMs to small perturbations \cite{Jaramillo:2020tuu,Jaramillo:2021tmt,Destounis:2021lum,Jaramillo:2022kuv,Cao:2024oud,Ianniccari:2024ysv,Boyanov:2022ark,Trefethen:2005}.

Environmental models introduce an additional difficulty. A one-parameter scan can show mode repulsion, but it does not generally satisfy the two real conditions required to locate an isolated EP \cite{Onozawa:1996ux,Cook:2014cta,Dias:2021yju,Dias:2022oqm,Davey:2023fin}. A small frequency gap is therefore only a locator, and a reliable calculation must combine eigenvector, left--right, Jordan-chain and closed-loop diagnostics. It must also distinguish continuum behavior from artifacts of finite spectral truncation. Astrophysical black holes are embedded in accretion flows and matter distributions, so environmental corrections can affect ringdown observables \cite{Barausse:2014tra,Cardoso:2021wlq,Zou:2020jeu,Zou:2020jsn}. We use the static, spherically symmetric string-fluid solution of Rastall gravity. Its lapse contains a dimensionless parameter $A$ and a sign $\eR=\pm1$ \cite{Rastall:1972swe,Letelier:1979,Barbosa:2016,Letelier:1981,Soleng:1995,Bezerra:2022srj}. The fluid contribution is a power law for $A\ne2$ and logarithmic for $A=2$, and the scalar QNM spectrum of this family has been studied previously \cite{Zhang:2024bov}, but its EP structure and associated Jordan response have not been examined.

To address these issues, we add a localized P\"oschl--Teller perturbation to the scalar potential and vary its position $a$ and amplitude $\eb$ as independent controls. We locate an isolated second-order EP by solving the two real equations defined by the complex discriminant. The QNM problem is compactified on hyperboloidal slices and discretized with two-domain Chebyshev methods \cite{Zenginoglu:2011jz,Ansorg:2016ztf,PanossoMacedo:2023qzp,PanossoMacedo:2024nkw}. Arbitrary-precision calculations, Jordan-chain diagnostics, closed-loop transport and constrained time evolution are used as complementary tests. We consider both the regular $A=1$ branch and the logarithmic $A=2$ branch, and require resolution convergence before interpreting a finite-dimensional degeneracy as a continuum EP. The bump is phenomenological and is not a solution of the Rastall field equations, so the calculation tests a controlled spectral mechanism rather than a universal astrophysical one. Motivated by these considerations, we investigate the EP structure and Jordan response of scalar QNMs in this controlled Rastall string-fluid environment.

Section~\ref{sec:model-method} defines the background, perturbation, compactification, and diagnostics. Sections~\ref{sec:a1-frequency} and \ref{sec:a1-time} present the $A=1$ frequency- and time-domain results. Sections~\ref{sec:a2-frequency} and \ref{sec:a2-time} describe the finite-dimensional $A=2$ results. Section~\ref{sec:conclusion} summarizes the conclusions. We use geometrized units with $G=c=1$.

\section{Model and numerical method}
\label{sec:model-method}

\subsection{Background geometry and scalar perturbations}

The static, spherically symmetric line element is
\begin{equation}
 \dd s^2=-f(r)\dd t^2+\frac{\dd r^2}{f(r)}
 +r^2(\dd\theta^2+\sin^2\theta\,\dd\phi^2).
 \label{eq:metric}
\end{equation}
For a black hole surrounded by a string fluid in Rastall gravity, the metric function is \cite{Bezerra:2022srj,Zhang:2024bov}
\begin{equation}
 f(r)=1-\frac{2M}{r}+\frac{Q^2}{r^2}-
 \begin{cases}
  \eR L r^{-1}\log(\lambda_{\mathrm R}r),&A=2,\\[2mm]
  \displaystyle \eR\frac{A}{A-2}
  \left(\frac{L}{r}\right)^{2/A},&A\ne2.
 \end{cases}
 \label{eq:lapse-general}
\end{equation}
Here $Q$ is the electric charge, $L$ and $\lambda_{\mathrm R}$ are positive parameters, and $\eR=\pm1$ fixes the sign of the string-fluid energy density. We use $\eR$ exclusively for the Rastall parameter and $\eb$ for the amplitude of the perturbing bump. For the $A=1$, $\eR=-1$ branch, Eq.~\eqref{eq:lapse-general} reduces to
\begin{equation}
 f(r)=1-\frac{2M}{r}+\frac{Q^2-L^2}{r^2}
 =\frac{(r-r_+)(r-r_-)}{r^2},
 \label{eq:lapse}
\end{equation}
where
\begin{equation}
 r_{\pm}=M\pm\sqrt{M^2+L^2-Q^2}.
 \label{eq:horizons}
\end{equation}
The outer event horizon is $r_h=r_+$ and the nonextremal condition is $Q^2<M^2+L^2$. For $M=L=1$, $Q_{\rm ext}=\sqrt2$; the special case $Q=1$ gives $f=1-2/r$ and provides a useful reduction test against the Schwarzschild code.

For a massless scalar field, $\Box\varphi=0$, we separate
\begin{equation}
 \varphi=\frac{\psi(r)}{r}Y_{\ell m}(\theta,\phi)e^{\ii\omega t}.
 \label{eq:separation}
\end{equation}
We use $e^{+\ii\omega t}$, so damped modes have $\operatorname{Im}\omega>0$. The usual $e^{-\ii\omega t}$ convention is obtained by complex conjugation. The oscillation frequency and damping rate are unchanged.
The radial equation is
\begin{equation}
 \frac{\dd^2\psi}{\dd r_*^2}+\left[\omega^2-V_s(r)\right]\psi=0,
 \qquad \frac{\dd r_*}{\dd r}=\frac1{f(r)},
 \label{eq:radial}
\end{equation}
with effective potential
\begin{equation}
 V_s(r)=f(r)\left[\frac{\ell(\ell+1)}{r^2}
 +\frac{f'(r)}r\right].
 \label{eq:scalar-potential}
\end{equation}
To create a tunable two-dimensional non-Hermitian parameter space, we add a P\"oschl--Teller-type bump,
\begin{equation}
 V_{\rm bump}(r_*)=\frac{\eb}{r_h^2}
 \sech^2\left(\frac{r_*}{r_h}-a\right).
 \label{eq:bump}
\end{equation}
The parameters $a$ and $\eb$ are dimensionless, and the physical bump amplitude is $\eb/r_h^2$. We use this term only as a phenomenological local environmental correction \cite{Cheung:2021bol,PanossoMacedo:2025xnf}.

For the convergent calculation, we use the regular $A=1$, $\eR=-1$ branch with $M=L=1$, $Q=1.2$, and $\ell=2$. The horizons are $r_-=0.2516685M$ and $r_h=1.7483315M$. Only $r>r_h$ enters the QNM problem. At the reported EP, the bump lies on the outgoing tail and reaches $0.93\%$ of the background barrier height.
\begin{figure}[htbp]
\centering
\includegraphics[width=0.98\textwidth]{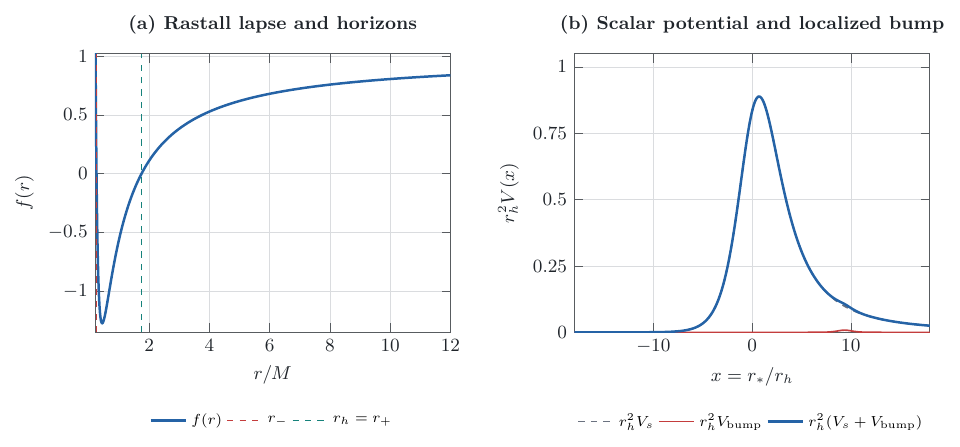}
\caption{Background geometry and localized perturbation for the regular $A=1$, $\eR=-1$ Rastall string-fluid branch with $M=L=1$, $Q=1.2$ and $\ell=2$. (a) The lapse $f(r)$, with dashed lines marking the inner root $r_-$ and the outer event horizon $r_h=r_+$. The physical QNM exterior is the positive-lapse region to the right of $r_h$. (b) The dimensionless scalar potential $r_h^2V_s$, the P\"oschl--Teller bump $r_h^2V_{\rm bump}$ and their sum as functions of $x=r_*/r_h$, evaluated at $a=9.3722477278$ and $\epsilon_{\mathrm b}=0.00822962356$. The bump is placed on the outgoing tail of the scalar-potential barrier.}
\label{fig:background-geometry}
\end{figure}

\subsection{Hyperboloidal spectral discretization}

The QNM equation is posed on the noncompact tortoise line $r_*\in(-\infty,+\infty)$. Its solutions are ingoing at the horizon and outgoing at infinity. We therefore use a hyperboloidal compactification \cite{Zenginoglu:2011jz,PanossoMacedo:2024nkw}. The compactified slices reach both characteristic boundaries, so regularity replaces artificial boundary conditions at finite radius.

For this purpose, we define
\begin{equation}
 \sigma=\frac{r_h}{r},\qquad x=\frac{r_*}{r_h},\qquad
 \kappa=\frac{r_-}{r_+}.
\end{equation}
Here $\sigma$ maps the complete black-hole exterior $r\in[r_h,\infty)$ to the finite interval $\sigma\in[0,1]$: $\sigma=1$ is the outer event horizon and $\sigma=0$ is future null infinity $\scri^+$. The coordinate $x$ measures the tortoise distance in units of $r_h$, so the bump position $a$ in Eq.~\eqref{eq:bump} is dimensionless, while $\kappa$ records the separation of the two roots of the lapse. In the nonextremal exterior considered below, $\kappa<1$. The metric function consequently becomes
\begin{equation}
 f(\sigma)=(1-\sigma)(1-\kappa\sigma),
\end{equation}
and the dimensionless tortoise coordinate has the analytic form
\begin{align}
 x(\sigma)={}&\frac1\sigma-(1+\kappa)\log\sigma
 +\frac{\log(1-\sigma)}{1-\kappa}\nonumber\\
 &-\frac{\kappa^2}{1-\kappa}\log(1-\kappa\sigma).
 \label{eq:tortoise}
\end{align}
As $\sigma\to0$, the $1/\sigma$ term sends $x$ to $+\infty$, whereas the $\log(1-\sigma)$ term sends $x$ to $-\infty$ as $\sigma\to1$. Thus, Eq.~\eqref{eq:tortoise} represents the full tortoise line on a finite computational interval without introducing an artificial outer boundary. It also fixes the additive constant of $r_*$ and hence the numerical origin of the control parameter $a$; all $A=1$ values of $a$ quoted below use this convention.

The compactification alone does not fix the time slicing. We use the minimal gauge that reduces to the Schwarzschild choice of Ref.~\cite{PanossoMacedo:2025xnf}. Its height function $H(\sigma)$ is given in the Appendix and satisfies $\gamma=pH'$. The resulting coefficients are
\begin{align}
 p(\sigma)&=\sigma^2(1-\sigma)(1-\kappa\sigma),\label{eq:p}\\
 \gamma(\sigma)&=1-
 \frac{2\sigma^2(1-\kappa\sigma)}{1-\kappa},\label{eq:gamma}\\
 w(\sigma)&=\frac{4[(1-\kappa)(1+\sigma)-\kappa\sigma^2]}
 {(1-\kappa)^2}.\label{eq:w}
\end{align}
The coefficient $p$ multiplies the compactified radial operator. The functions $\gamma$ and $w$ set the slice tilt and time-derivative normalization. Since $p$ vanishes at both endpoints, while $\gamma(0)=1$ and $\gamma(1)=-1$, the endpoints are characteristic. For $Q=1$, hence $\kappa=0$, the coefficients reduce to the Schwarzschild expressions used as a code check.

The scalar potential in Eq.~\eqref{eq:scalar-potential} gives the polynomial spectral coefficient
\begin{equation}
 q_s(\sigma)=\ell(\ell+1)+
 \sigma(1+\kappa-2\kappa\sigma).
 \label{eq:qscalar}
\end{equation}
Including the bump, the total coefficient is
\begin{equation}
 q(\sigma)=q_s(\sigma)+\eb\,
 \frac{\sech^2[x(\sigma)-a]}{p(\sigma)}.
 \label{eq:qtotal}
\end{equation}
Although Eq.~\eqref{eq:qtotal} contains $1/p$, the factor $\sech^2[x-a]$ decays faster than $p$ at both endpoints. The bump therefore vanishes at $\sigma=0$ and $1$. We evaluate it in exponential form to avoid overflow for large $|x-a|$.
The first-order spatial operators are
\begin{equation}
 \mathcal L_1=p\,\partial_\sigma^2+p'\partial_\sigma-q,
 \qquad
 \mathcal L_2=2\gamma\partial_\sigma+\gamma'.
\end{equation}
After division by the positive weight $w$ and spectral discretization, they give the generalized eigenvalue problem
\begin{equation}
 \bm L\bm u=s\bm B\bm u,
 \qquad s=\ii\omega,
 \label{eq:gevp}
\end{equation}
consistent with the convention fixed by Eq.~\eqref{eq:separation}, so that a damped mode has $\operatorname{Re}s<0$.
We split the interval at the bump centre $x(\sigma)=a$ and use two Chebyshev--Lobatto subdomains. Each subdomain has $N+1$ points. The field and its first derivative are matched at the interface. We assemble the $N$ and $N-5$ systems together and retain only damped modes that match between resolutions. This removes migrating pseudospectral points from mode continuation.

The two-dimensional search uses Float64 arithmetic. We then rebuild the complete problem at prescribed MPFR precision because defective eigenvalues are intrinsically ill conditioned. Horizon radii, interface locations, coordinates, Chebyshev matrices, potentials, and the pencil $(\bm L,\bm B)$ are recomputed from decimal strings. The interface constraints make $\bm B$ singular, so we use the shift-invert matrix
\begin{equation}
 {\bm M}_{s_0}=({\bm L}-s_0{\bm B})^{-1}{\bm B},
 \qquad s=s_0+\theta^{-1},
 \label{eq:shiftinvert}
\end{equation}
Here $\theta$ is an eigenvalue of ${\bm M}_{s_0}$ and $s_0=0$. Factorizations, Schur iterations, and singular-value decompositions use arbitrary-precision complex arithmetic. We use 192 bits for $N\le42$ and 128 bits for $N\ge48$. At $N=30$, both precisions agree in the centre frequency to about 37 decimal digits. The residuals, $1.7\times10^{-37}$ and $1.0\times10^{-56}$, are well below the reported truncation error.

\subsection{EP localization and numerical diagnostics}

A one-dimensional scan cannot distinguish an EP from an avoided crossing. Finite spectral pencils can also contain migrating pseudospectral points. We therefore fix $Q$ and $\ell$ and use the frequency gap only as a locator \cite{Kato1995,Heiss:1999alz,Heiss:2012dx,Trefethen:2005,Jaramillo:2020tuu,Destounis:2021lum,PanossoMacedo:2025xnf}.
\begin{equation}
 g(a,\eb)=|\omega_1-\omega_2|
 \label{eq:gap}
\end{equation}
We first select the six lowest damped modes that match between the principal and reference resolutions. The interacting pair is tracked by continuity, not by damping-rate sorting. We refine the local gap map around this pair. Direct gap minimization is poorly conditioned because $\Delta\omega=\omega_1-\omega_2$ has a square-root branch point \cite{Heiss:2012dx,PanossoMacedo:2025xnf}. We therefore use the discriminant
\begin{equation}
 D(a,\eb)=[\Delta\omega(a,\eb)]^2
 \label{eq:discriminant}
\end{equation}
The discriminant removes the local square-root branch. We solve $\operatorname{Re}D=\operatorname{Im}D=0$ with a two-dimensional Newton iteration and a finite-difference Jacobian. These equations implement the codimension-two condition for an isolated second-order EP \cite{Kato1995,Heiss:2012dx,Cavalcante:2024swt}.

The residual gap is controlled by the Newton tolerance and by eigenvalue conditioning. It is not, by itself, evidence for a degeneracy. Higher precision lowers the round-off floor at fixed $N$, but it does not prove eigenvector coalescence \cite{Trefethen:2005,Jaramillo:2020tuu}. We therefore use four complementary diagnostics. The first is the phase-invariant eigenvector overlap
\begin{equation}
 C_R=\frac{|\bm v_1^\dagger\bm v_2|}
 {\|\bm v_1\|\,\|\bm v_2\|},
 \label{eq:overlap}
\end{equation}
which tends to unity at a second-order EP but remains bounded away from unity at an avoided crossing \cite{Kato1995,Heiss:2012dx}. The second diagnostic is the self-orthogonality measure
\begin{equation}
 R_j=\frac{|\bm y_j^\dagger\bm B\bm v_j|}
 {\|\bm y_j\|\,\|\bm B\bm v_j\|},
 \label{eq:selforth}
\end{equation}
At a defective degeneracy, the left and right eigenvectors become mutually orthogonal and $R_j$ vanishes \cite{Heiss:2012dx}. We also record the smallest singular value of $\bm L-s_c\bm B$, normalized by the matrix norm. The two quantities test the near-nullspace and the left--right structure of the pencil together.

The third diagnostic is an explicit Jordan chain. It tests defectiveness directly rather than through a proxy \cite{Kato1995,Heiss:2012dx}. With ${\bm K}_c={\bm L}-s_c{\bm B}$, the eigenvector ${\bm u}_0$ and generalized eigenvector ${\bm u}_1$ satisfy
\begin{equation}
 {\bm K}_c{\bm u}_0=0,\qquad
 {\bm K}_c{\bm u}_1={\bm B}{\bm u}_0.
 \label{eq:jordan-chain}
\end{equation}
We take the smallest right singular vector of ${\bm K}_c$ as ${\bm u}_0$. We construct ${\bm u}_1$ with an SVD pseudoinverse after removing that singular direction. With $S=\|{\bm L}\|+|s_c|\|{\bm B}\|$, the residuals
\begin{align}
 r_0&=\frac{\|{\bm K}_c{\bm u}_0\|}{S\|{\bm u}_0\|},\label{eq:r0}\\
 r_1&=\frac{\|{\bm K}_c{\bm u}_1-{\bm B}{\bm u}_0\|}
 {S\|{\bm u}_1\|+\|{\bm B}{\bm u}_0\|}
 \label{eq:r1}
\end{align}
measure the two chain equations. The second equation is solvable only when ${\bm B}{\bm u}_0$ lies in the range of ${\bm K}_c$. For the smallest left singular vector ${\bm y}_0$, this requires the compatibility measure
\begin{equation}
 \chi_0=\frac{|{\bm y}_0^\dagger{\bm B}{\bm u}_0|}
 {S\|{\bm y}_0\|\|{\bm B}{\bm u}_0\|}
 \label{eq:chi0}
\end{equation}
At an EP, $\chi_0$ vanishes. Together, $r_0$, $r_1$, and $\chi_0$ establish a length-two chain; eigenfunction overlap alone cannot do so.

The fourth diagnostic is topological. Because Eq.~\eqref{eq:sqrt} has a square-root branch point, continuous transport around an enclosing loop must exchange the two frequencies, $(\omega_1,\omega_2)\mapsto(\omega_2,\omega_1)$. The original assignment returns after a second circuit \cite{Heiss:2012dx,Bergholtz:2019deh,Cavalcante:2024swt}. This test probes a different property from local conditioning: it requires continuous tracking and an enclosing loop, not merely a small gap at one point. We evaluate every diagnostic at independently relocated candidates for each $N$, which allows refinement stability to be tested.

We distinguish three related statements. A \emph{finite-dimensional EP} is a defective degeneracy of the discretized pencil $(\bm L,\bm B)$ at fixed $N$. A \emph{continuum EP} also requires the relocated parameters and frequencies to converge as $N$ increases. We call a finite-dimensional EP \emph{numerical} when this refinement stability is demonstrated. The $A=1$ results meet this criterion; the $A=2$ results do not.

\section{\texorpdfstring{$A=1$ branch: frequency-domain results}{A=1 branch: frequency-domain results}}
\label{sec:a1-frequency}

\subsection{Search setup and EP location}

For the $A=1$, $\eR=-1$ branch we fix
\begin{equation}
 M=L=1,\qquad Q=1.2,\qquad \ell=2,
 \label{eq:a1-parameters}
\end{equation}
The charge is below the extremal value $Q_{\rm ext}=\sqrt2$, and $\kappa=r_-/r_+$ is nonzero. An initial scan over $a\in[2,18]$ and $\eb\in[5\times10^{-4},8\times10^{-3}]$ placed the smallest gap on the upper amplitude boundary. We then enlarged and narrowed the search to $a\in[8.9,9.8]$ and $\eb\in[0.0075,0.0115]$, where the minimum moved into the interior.

\begin{figure}[htbp]
\centering
\includegraphics[width=0.98\textwidth]{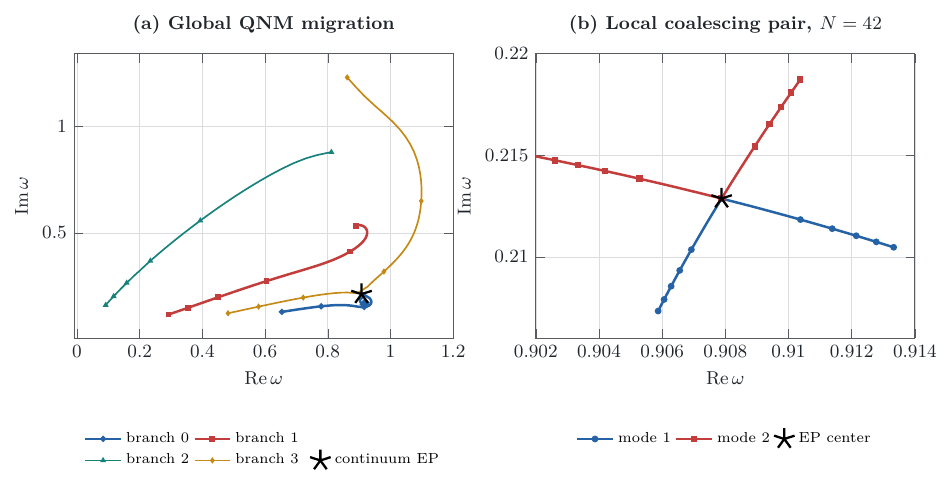}
\caption{Evolution of the QNMs of the $A=1$ branch in the complex-frequency plane, with fixed $M=L=1$, $Q=1.2$, $\eR=-1$ and $\ell=2$. The left panel tracks four low-lying branches at $N=30$, while the right panel resolves the local coalescing pair near the EP at $N=42$. The star marks the final reported EP center frequency, not an $N=30$ relocation.}
\label{fig:qnm-migration}
\end{figure}

Figure~\ref{fig:qnm-migration} tracks four low-lying branches at $N=30$ while $a$ varies at fixed amplitude. Labels follow continuity between neighboring points. Two branches approach $\omega\simeq0.908+0.213\ii$ and form the interacting pair. The other two remain separated. The coarse step $\Delta a=0.2$ misses the rapid motion near the degeneracy, so we refine the search locally. This scan identifies the pair but cannot locate the EP without the second control parameter.

\begin{figure}[htbp]
\centering
\includegraphics[width=0.82\textwidth]{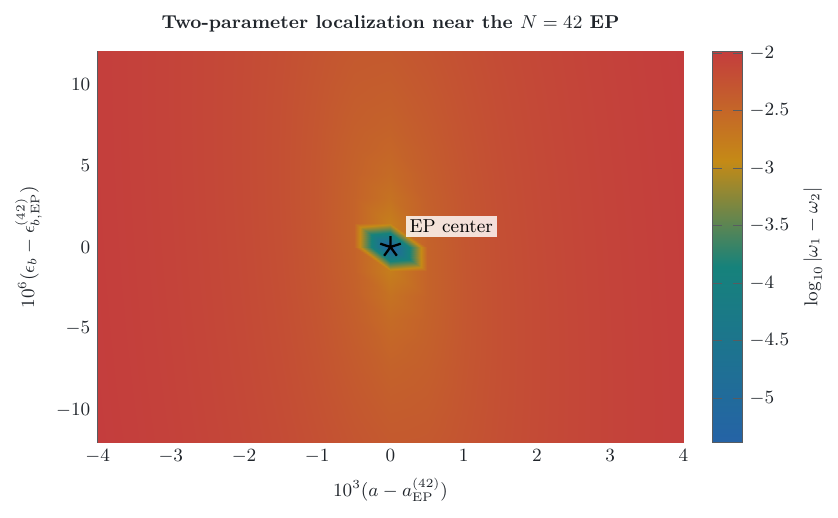}
\caption{Two-parameter localization near the $N=42$ EP of the $A=1$ branch, with fixed $M=L=1$, $Q=1.2$, $\eR=-1$ and $\ell=2$. The color represents $\log_{10}|\omega_1-\omega_2|$ in the $(a,\eb)$ plane; the axes are scaled by $10^3$ and $10^6$, respectively. The star marks the EP center relocated with arbitrary-precision arithmetic, and the isolated minimum identifies the interacting pair.}
\label{fig:gap-map}
\end{figure}

We centre a $17\times17$ grid on the independently relocated $N=42$ candidate, with $|a-a_{\rm EP}^{(42)}|\le0.004$ and $|\eb-\epsilon_{{\rm b},{\rm EP}}^{(42)}|\le1.2\times10^{-5}$. Figure~\ref{fig:gap-map} shows $\log_{10}|\omega_1-\omega_2|$ in this box. The gap decreases towards an isolated minimum rather than an extended valley. The nearest double-precision point has a gap of $4.107\times10^{-6}$. We obtain the reported coordinates by solving $\operatorname{Re}D=\operatorname{Im}D=0$ at arbitrary precision.

\FloatBarrier

For each $N$, we solve $\operatorname{Re}D=\operatorname{Im}D=0$ independently and seed Newton's iteration with the previous resolution. Table~\ref{tab:location} lists the resulting relocations. The final-column gap $g$ records the stopping tolerance; it is not a convergence measure. We assess the continuum limit from the changes in $a_N$, $\eb^{(N)}$, and the centre frequency.

\begin{table}[htbp]
\centering
\caption{Arbitrary-precision EP relocations of the $A=1$ branch at different spectral resolutions, with fixed $M=L=1$, $Q=1.2$, $\eR=-1$ and $\ell=2$. The bit column gives the working precision and $g=|\omega_1-\omega_2|$ is the residual gap at the end of the Newton iteration. The stabilized coordinates demonstrate resolution convergence.}
\label{tab:location}
\small
\begin{tabular}{rrrrr}
\toprule
$N$ & bit & $a_N$ & $\eb^{(N)}$ & $g$\\
\midrule
24 & 192 & 9.372092364670 & 0.008219215956293 & $3.80\times10^{-15}$\\
30 & 192 & 9.372241945843 & 0.008229843671315 & $3.94\times10^{-15}$\\
36 & 192 & 9.372247930435 & 0.008229619991626 & $4.16\times10^{-17}$\\
42 & 192 & 9.372247724070 & 0.008229623577239 & $3.49\times10^{-14}$\\
48 & 128 & 9.372247727635 & 0.008229623556639 & $1.21\times10^{-11}$\\
54 & 128 & 9.372247727790 & 0.008229623557194 & $2.57\times10^{-12}$\\
60 & 128 & 9.372247727781 & 0.008229623557401 & $7.05\times10^{-10}$\\
\bottomrule
\end{tabular}
\end{table}

From $N=42$ to $60$, the changes in $a_N$, $\eb^{(N)}$, and the centre frequency decrease by several orders of magnitude. The final three refinements are $3.56\times10^{-9}$, $1.55\times10^{-10}$, and $8.89\times10^{-12}$ for $a_N$. The corresponding changes in $\eb^{(N)}$ are $2.06\times10^{-11}$, $5.54\times10^{-13}$, and $2.07\times10^{-13}$. We use the last difference as a conservative truncation estimate.
\begin{align}
 a_{\rm EP}&=9.372247727781(9),\label{eq:aep}\\
 \eb^{\rm EP}&=0.0082296235574(3),\label{eq:epsep}\\
 \omega_{\rm EP}&=0.907881134084(2)+0.212872971497(1)\,\ii.\label{eq:omegaep}
\end{align}
The parenthetical digits are truncation estimates, not confidence intervals. An Aitken $\Delta^2$ extrapolation of $N=48,54,60$ agrees with the reported values. Because the spectral error is not strictly monotonic, we use the extrapolation only as a consistency check.

\subsection{Defectiveness and the Jordan chain}

Frequency coalescence can occur at both defective and diabolical degeneracies. Table~\ref{tab:diagnostics} therefore reports the eigenvector overlap, self-orthogonality, singular-value, and Jordan-chain tests at every relocated point. These diagnostics remain many orders of magnitude below unity for $N=36$--$60$. Their nonmonotonic variation follows the Newton tolerances and the increasing matrix conditioning. The eigen-equation residual remains below $1.6\times10^{-56}$ at 192 bits and $1.1\times10^{-36}$ at 128 bits.

\begin{table}[htbp]
\centering
\caption{Defectiveness and Jordan-chain diagnostics of the $A=1$ branch at arbitrary-precision relocations, with fixed $M=L=1$, $Q=1.2$, $\eR=-1$ and $\ell=2$. The diagnostics are $1-C_R$, the left--right self-orthogonality measure $R$, the normalized smallest singular value $\widehat\sigma_{\min}$, and the Jordan-chain residuals $r_0$, $r_1$, and $\chi_0$. The nonmonotonic dependence on $N$ reflects the differing Newton stopping tolerances and the growth of the matrix condition number near the degeneracy, not a loss of arithmetic accuracy.}
\label{tab:diagnostics}
\small
\resizebox{\textwidth}{!}{%
\begin{tabular}{rrrrrrrr}
\toprule
$N$ & $g$ & $1-C_R$ & $R$ & $\widehat\sigma_{\min}$ & $r_0$ & $r_1$ & $\chi_0$\\
\midrule
36 & $4.16\times10^{-17}$ & $4.06\times10^{-33}$ & $5.83\times10^{-36}$ & $2.28\times10^{-43}$ & $2.28\times10^{-43}$ & $3.46\times10^{-42}$ & $7.54\times10^{-42}$\\
42 & $3.49\times10^{-14}$ & $2.86\times10^{-27}$ & $2.13\times10^{-30}$ & $4.13\times10^{-38}$ & $4.13\times10^{-38}$ & $6.85\times10^{-37}$ & $1.49\times10^{-36}$\\
48 & $1.21\times10^{-11}$ & $3.44\times10^{-22}$ & $1.38\times10^{-25}$ & $1.45\times10^{-33}$ & $1.45\times10^{-33}$ & $2.60\times10^{-32}$ & $5.64\times10^{-32}$\\
54 & $2.57\times10^{-12}$ & $1.55\times10^{-23}$ & $3.41\times10^{-27}$ & $2.09\times10^{-35}$ & $2.09\times10^{-35}$ & $4.02\times10^{-34}$ & $8.71\times10^{-34}$\\
60 & $7.05\times10^{-10}$ & $1.16\times10^{-18}$ & $1.43\times10^{-22}$ & $5.41\times10^{-31}$ & $5.41\times10^{-31}$ & $1.11\times10^{-29}$ & $2.41\times10^{-29}$\\
\bottomrule
\end{tabular}}
\end{table}

At $N=54$, $r_0=2.09\times10^{-35}$, $r_1=4.02\times10^{-34}$, and $\chi_0=8.71\times10^{-34}$. The overlap and self-orthogonality measures are $1-C_R=1.55\times10^{-23}$ and $R=3.41\times10^{-27}$. The two eigen-directions have merged, while a generalized eigenvector satisfying Eq.~\eqref{eq:jordan-chain} remains. The discrete pencil therefore contains a length-two Jordan chain.

\begin{figure}[H]
\centering
\begin{tabular}{@{}cc@{}}
\begin{minipage}[t]{0.46\textwidth}
\centering
\includegraphics[width=\linewidth]{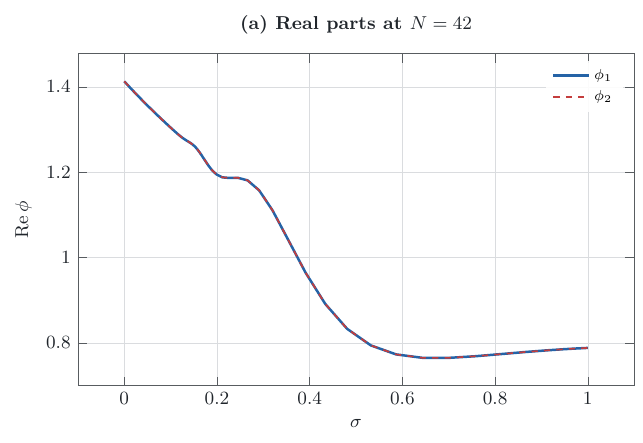}
\end{minipage} &
\begin{minipage}[t]{0.46\textwidth}
\centering
\includegraphics[width=\linewidth]{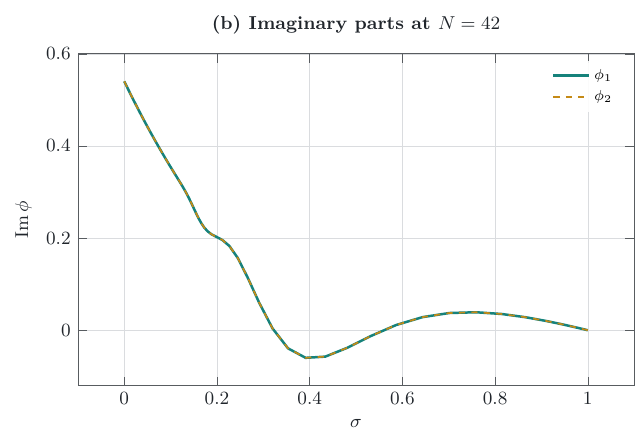}
\end{minipage}\\[0.7em]
\multicolumn{2}{c}{\begin{minipage}[t]{0.72\textwidth}
\centering
\includegraphics[width=\linewidth]{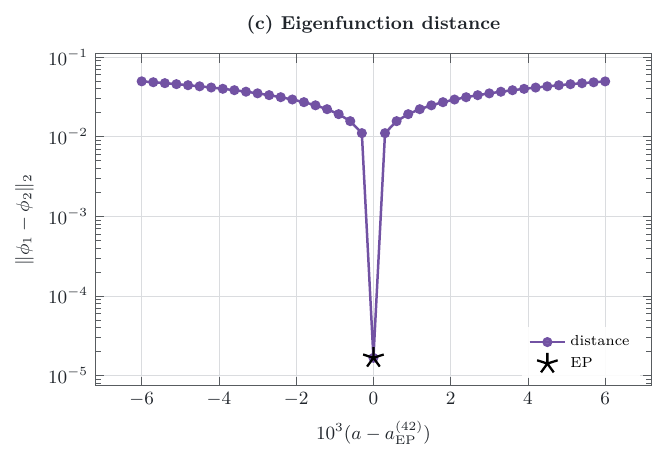}
\end{minipage}}
\end{tabular}
\caption{Radial-eigenfunction coalescence at the $N=42$ candidate of the $A=1$ branch, with fixed $M=L=1$, $Q=1.2$, $\eR=-1$ and $\ell=2$. (a) Real parts and (b) imaginary parts of the two unit-$L^2$-normalized, phase-aligned eigenfunctions; the solid and dashed curves denote $\phi_1$ and $\phi_2$, respectively. (c) Their distance at fixed $\eb=\epsilon_{{\rm b},{\rm EP}}^{(42)}$; the star marks the EP and the dashed line indicates the distance $1.660\times10^{-5}$ at that point.}
\label{fig:eigenfunction-coalescence}
\end{figure}

Figure~\ref{fig:eigenfunction-coalescence}(a)--(c) shows the same behavior in the radial eigenfunctions. The $N=42$ eigenfunctions are phase aligned and normalized to unit $L^2$ norm. Their distance reaches $1.660\times10^{-5}$ at the candidate, compared with a frequency gap of $4.107\times10^{-6}$. These curves use double precision; the quantitative defectiveness claim rests on the arbitrary-precision diagnostics in Table~\ref{tab:diagnostics}.

\begin{figure}[H]
\centering
\begin{tabular}{@{}cc@{}}
\begin{minipage}[t]{0.46\textwidth}
\centering
\includegraphics[width=\linewidth]{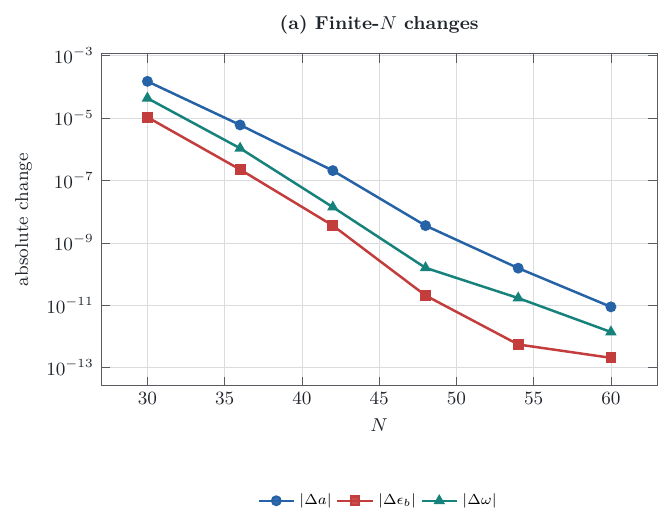}
\end{minipage} &
\begin{minipage}[t]{0.46\textwidth}
\centering
\includegraphics[width=\linewidth]{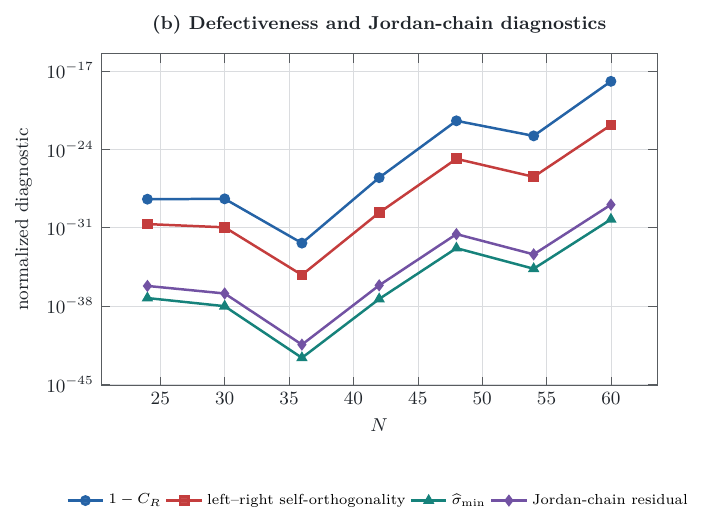}
\end{minipage}\\[0.7em]
\multicolumn{2}{c}{\begin{minipage}[t]{0.72\textwidth}
\centering
\includegraphics[width=\linewidth]{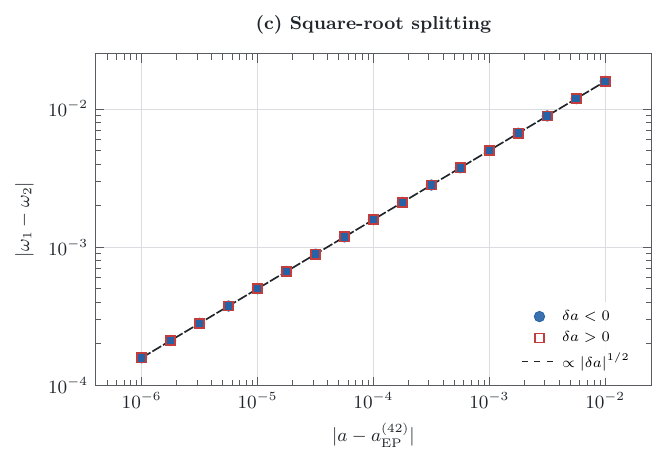}
\end{minipage}}
\end{tabular}
\caption{Resolution convergence and EP diagnostics for the $A=1$ branch, with fixed $M=L=1$, $Q=1.2$, $\eR=-1$ and $\ell=2$. (a) Adjacent-resolution changes in the arbitrary-precision relocation sequence: $|\Delta a|$, $|\Delta\eb|$, and $|\Delta\omega|$. (b) Defectiveness indicators: $1-C_R$, the left--right self-orthogonality measure $R$, the normalized smallest singular value $\widehat\sigma_{\min}$, and the Jordan-chain residual. (c) Frequency gap on the two sides of the $N=42$ EP; the solid curve is the numerical result and the dashed curve is the $|a-a_{\rm EP}^{(42)}|^{1/2}$ reference.}
\label{fig:convergence-jordan-sqrt}
\end{figure}

Figure~\ref{fig:convergence-jordan-sqrt}(a)--(c) summarizes the resolution study. From $N=30$ to $60$, the changes in $a$, $\eb$, and $\omega$ decrease by several orders of magnitude. The four defectiveness indicators remain small. At fixed $\eb=\epsilon_{{\rm b},{\rm EP}}^{(42)}$, the local splitting follows $|a-a_{\rm EP}^{(42)}|^{1/2}$. A logarithmic fit gives a slope of $0.4999999$ on both sides of the candidate, consistent with the Puiseux form in Eq.~\eqref{eq:sqrt}.

\FloatBarrier

\subsection{Closed-loop transport and mode exchange}

\begin{figure}[htbp]
\centering
\includegraphics[width=0.98\textwidth]{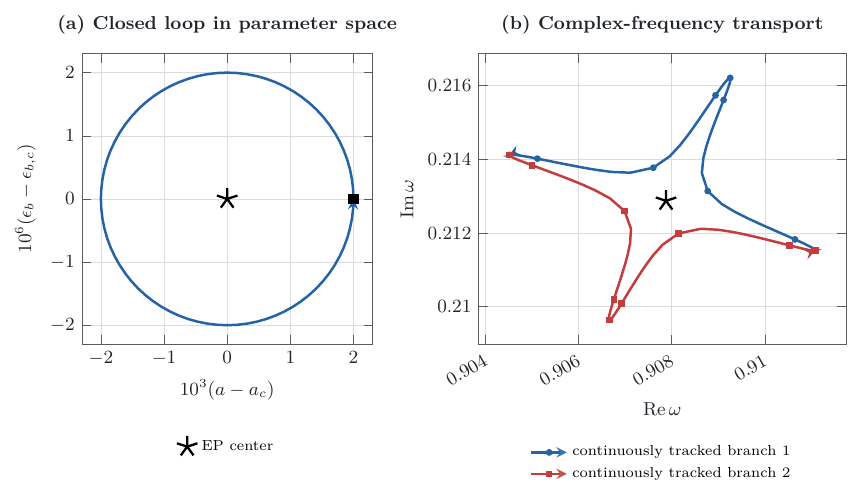}
\caption{Closed-loop transport of the $A=1$ branch at $N=54$, with fixed $M=L=1$, $Q=1.2$, $\eR=-1$ and $\ell=2$. (a) The loop in the $(a,\eb)$ plane. (b) The continuously tracked complex-frequency trajectories. Arrows indicate the transport direction, and the star marks the EP center. One circuit exchanges the two modes.}
\label{fig:mode-exchange}
\end{figure}

We next test the topology of the candidate. At $N=54$, we continue the two modes around the closed parameter loop
\begin{equation}
 \begin{split}
  a(\theta)&=a_0+0.002\cos\theta,\\
  \eb(\theta)&=\epsilon_{{\rm b},0}+2\times10^{-6}\sin\theta,
 \end{split}
 \qquad 0\le\theta\le2\pi,
 \label{eq:loop}
\end{equation}
using 80 equally spaced points. The loop is centred at
\[
(a_0,\epsilon_{{\rm b},0})=(9.3722477276,\ 0.00822962356).
\]
This centre differs from the converged coordinates by $1.8\times10^{-10}$ in $a$ and $2.6\times10^{-12}$ in $\eb$, much less than the loop semiaxes. The branches are tracked by frequency continuity. Figure~\ref{fig:mode-exchange} shows the parameter loop and the corresponding frequency trajectories.
At $\theta=0$, the two branches are
$\omega_1(0)=0.9110853+0.2115250\,\ii$ and
$\omega_2(0)=0.9045176+0.2141175\,\ii$. After one circuit, branch~1 ends at $\omega_2(0)$ and branch~2 ends at $\omega_1(0)$. The distances for the direct and swapped assignments are
\begin{equation}
 d_{\rm direct}=1.412163216\times10^{-2},
 \qquad d_{\rm swapped}=0,
 \label{eq:a1-loop-result}
\end{equation}
The swapped distance is zero to double precision. The branches therefore exchange after one circuit and return after two, as expected for a square-root branch point. Together with the local diagnostics and the Jordan chain, this identifies a second-order numerical EP. We use one representative loop at fixed resolution. Loop-radius and sampling-density convergence are not tested here.

\section{\texorpdfstring{$A=1$ branch: time-domain response}{A=1 branch: time-domain response}}
\label{sec:a1-time}

\subsection{Constraint elimination and a stable initial-value problem}

The interface conditions replace two rows of the discrete system. Thus $\bm B$ is singular and cannot be inverted directly. We eliminate the constraints explicitly. Let $\bm\Phi$ and $\bm\Psi$ denote the field and its time derivative on the two subdomains. With an embedding matrix $\bm Z$ satisfying $\bm C\bm Z=0$, we set
\begin{equation}
 \bm\Phi=\bm Z\bm q,\qquad \bm\Psi=\bm Z\bm p
 \label{eq:embedding}
\end{equation}
This parameterizes the constraint-satisfying subspace by $(\bm q,\bm p)$. Projection onto the remaining dynamical rows gives
\begin{equation}
 \frac{\dd}{\dd t}
 \begin{pmatrix}\bm q\\ \bm p\end{pmatrix}
 =\bm A_{\rm red}
 \begin{pmatrix}\bm q\\ \bm p\end{pmatrix},\qquad
 \bm A_{\rm red}=\begin{pmatrix}\bm0&\bm I\\ \bm A_1&\bm A_2\end{pmatrix}.
\label{eq:reduced-ivp}
\end{equation}
We sample the waveform uniformly with $\Delta t=0.05$. The propagation operator is the matrix exponential
\begin{equation}
 \bm P=\exp(\bm A_{\rm red}\Delta t),
\end{equation}
It is applied repeatedly, so $\Delta t$ is a sampling interval rather than an ODE integration step. We use static initial data $(\Phi,\partial_t\Phi)|_{t=0}=(1,0)$ and read the waveform at $\scri^+$. At $N=42$, the constraint residual is $5.1\times10^{-20}$ and the normalized residual in the original equations is $8.8\times10^{-18}$.

\begin{figure}[htbp]
\centering
\includegraphics[width=0.98\textwidth]{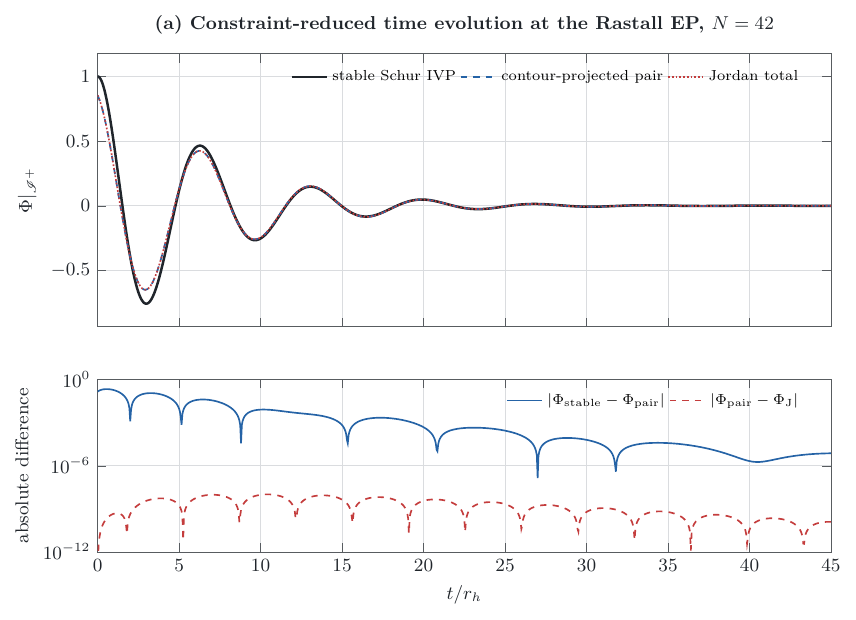}
\caption{Time-domain waveform of the $A=1$ branch at $N=42$, with fixed $M=L=1$, $Q=1.2$, $\eR=-1$, $\ell=2$ and the EP coordinates in Eqs.~\eqref{eq:aep}--\eqref{eq:omegaep}. The black solid curve is the initial-value evolution in the stable Schur subspace, the blue dashed curve is the EP two-dimensional contribution, and the red dotted curve is the Jordan reconstruction; the lower panel shows the absolute differences from the EP-pair propagation.}
\label{fig:td-waveform}
\end{figure}

The generator also contains spurious growing eigenvalues. At $N=42$, the largest has $\operatorname{Re}s=3424$. Direct matrix exponentiation would amplify round-off faster than the physical signal decays. We compute an ordered complex Schur decomposition and retain the invariant subspace with $\operatorname{Re}s<0$. The evolution is
\begin{equation}
 \bm x_{\rm st}(t)=\bm Q_{\rm st}
 \exp(\bm T_{\rm st}t)\bm Q_{\rm st}^{\dagger}\bm x_0,
 \qquad \bm T_{\rm st}=\bm Q_{\rm st}^{\dagger}
 \bm A_{\rm red}\bm Q_{\rm st}.
 \label{eq:stable-ivp}
\end{equation}
This procedure discards the growing directions. At $N=42$, the projection changes the initial state by $1.05\times10^{-9}$ and has an invariance residual of $4.61\times10^{-13}$. Figure~\ref{fig:td-waveform} compares the stable evolution, the EP pair, and its Jordan representation.

\subsection{Riesz projection and the linear-in-time term}

To isolate the EP contribution without fitting it, we choose a contour $\Gamma$ in the complex $s$ plane that encloses the interacting pair and no other eigenvalue. We compute the Riesz projector
\begin{equation}
 \bm P_{\rm EP}=\frac{1}{2\pi\ii}
 \oint_\Gamma(z\bm I-\bm A_{\rm red})^{-1}\dd z.
 \label{eq:riesz-projector}
\end{equation}
We evaluate the integral with a 96-point periodic trapezoidal rule. The first two left singular vectors of $\bm P_{\rm EP}$ define the EP subspace. At $N=42$, the third-to-second singular-value ratio is $5.7\times10^{-11}$ and the invariant-subspace residual is $2.47\times10^{-9}$. No contour-density refinement is claimed.
We define
\begin{equation}
 \bm T_{\rm EP}=\bm Q_{\rm EP}^{\dagger}\bm A_{\rm red}\bm Q_{\rm EP},
 \quad \bm c=\bm Q_{\rm EP}^{\dagger}\bm P_{\rm EP}\bm x_{\rm st}(0),
 \quad \bm N=\bm T_{\rm EP}-s_c\bm I,
 \label{eq:ep-compression}
\end{equation}
Here $s_c$ is the mean of the enclosed eigenvalues, $\bm c$ is the projected initial state, and $\bm h^T$ is the observation functional at $\scri^+$. The Jordan coefficients are
\begin{equation}
 C=\bm h^T\bm Q_{\rm EP}\bm c,
 \qquad D=\bm h^T\bm Q_{\rm EP}\bm N\bm c.
 \label{eq:jordan-cd}
\end{equation}
The corresponding real waveform is
\begin{equation}
 \Phi_{\rm J}(t)=2\operatorname{Re}\!\left[(C+Dt)e^{s_ct}\right]
 =2\operatorname{Re}(Ce^{s_ct})
 +2\operatorname{Re}(Dt e^{s_ct}).
 \label{eq:jordan-waveform}
\end{equation}
With $s=\ii\omega$, the second term is the expected $te^{+\ii\omega_{\rm EP}t}$ contribution of a second-order pole \cite{Yang:2025dbn,PanossoMacedo:2025xnf}. It comes from the nilpotent part $\bm N$ of the compressed generator, not from a fitting ansatz. At $N=42$,
\begin{equation}
 C=0.4270371+0.0990733\,\ii,\qquad
 D=0.0483232+0.0413018\,\ii.
 \label{eq:a1-jordan-cd-values}
\end{equation}
The compression satisfies $\|\bm N^2\bm c\|/\|\bm c\|=1.63\times10^{-9}$, so no $t^2$ term is resolved. Over $12\le t/r_h\le35$, the Jordan form agrees with $\exp(\bm T_{\rm EP}t)$ to $6.87\times10^{-8}$. The EP pair differs from the full stable waveform by $2.97\times10^{-2}$ in the same window.

\begin{figure}[htbp]
\centering
\includegraphics[width=0.96\textwidth]{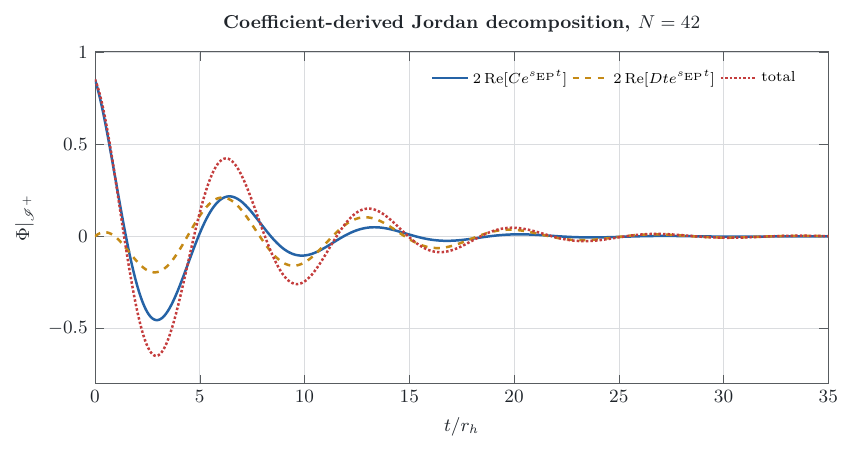}
\caption{Jordan response of the $A=1$ branch at $N=42$, with fixed $M=L=1$, $Q=1.2$, $\eR=-1$ and $\ell=2$. The curves show the constant term $2\operatorname{Re}(Ce^{s_ct})$, the linear term $2\operatorname{Re}(Dte^{s_ct})$, and their sum, which contains the characteristic linear-in-time contribution.}
\label{fig:jordan-components}
\end{figure}

\subsection{Comparison of three waveform models}

The linear term was obtained without fitting. We next compare three waveform parameterizations on the same Riesz-projected EP signal. Writing $s_i=\alpha_i+\ii\beta_i$ and $s_c=\alpha_c+\ii\beta_c$, we use
\begin{align}
 \texttt{2QNM}:&\quad 2\operatorname{Re}\!\left[\sum_{i=1}^{2}C_i e^{s_i t}\right],
 \qquad C_i,s_i\in\mathbb C, \\
 \texttt{EP linear}:&\quad 2\operatorname{Re}\!\left[D\,t e^{s_c t}\right],
 \qquad D,s_c\in\mathbb C, \\
 \texttt{EP total}:&\quad 2\operatorname{Re}\!\left[(C+Dt)e^{s_c t}\right],
 \qquad C,D,s_c\in\mathbb C.
 \label{eq:fit-models}
\end{align}
The three models contain 8, 4, and 6 free real parameters, respectively. At each trial frequency, the complex amplitudes are obtained by real linear least squares. The frequencies are optimized by bounded variable projection. All fits use $12\le t/r_h\le35$.

\begin{figure}[htbp]
\centering
\includegraphics[width=0.98\textwidth]{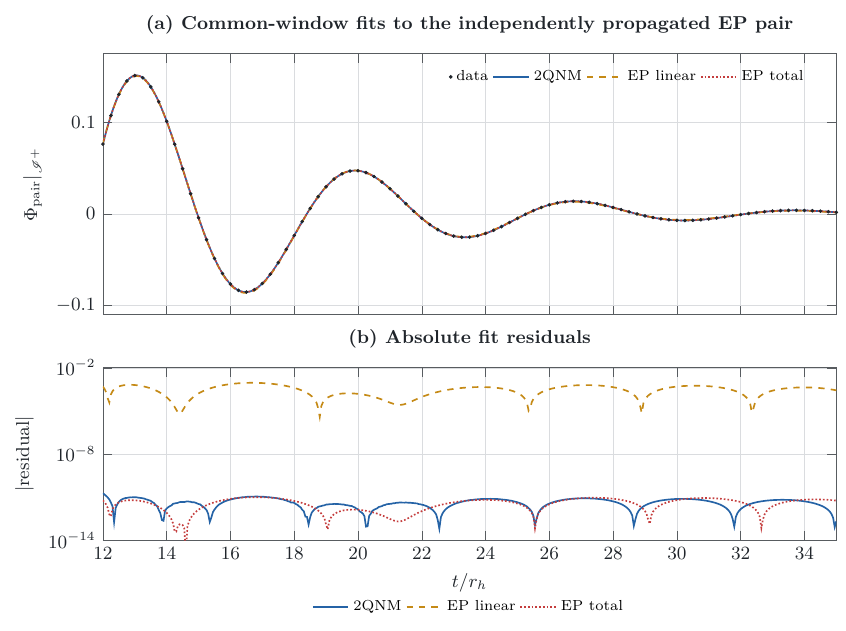}
\caption{Fits of the same $A=1$ EP two-mode signal with \texttt{2QNM}, \texttt{EP linear} and \texttt{EP total}, for fixed $M=L=1$, $Q=1.2$, $\eR=-1$, $\ell=2$ and $12\le t/r_h\le35$. The upper panel shows the waveforms and the lower panel the absolute residuals.}
\label{fig:three-fits}
\end{figure}

\begin{figure}[htbp]
\centering
\includegraphics[width=0.98\textwidth]{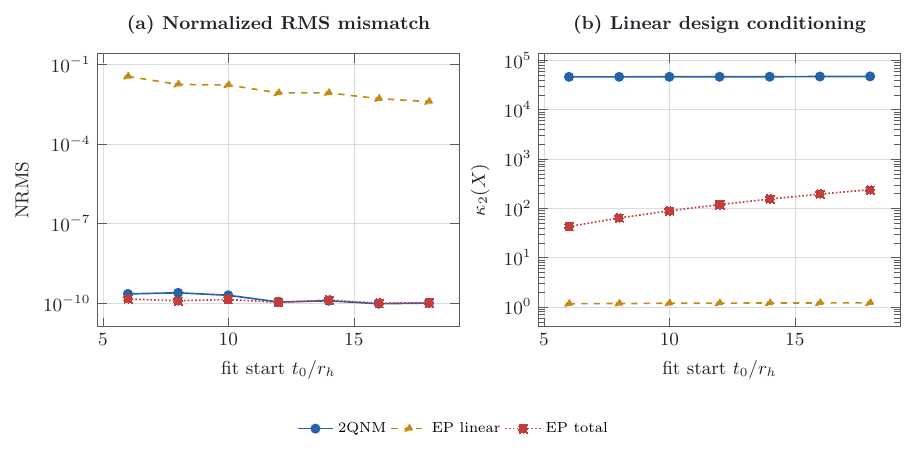}
\caption{Dependence of the three $A=1$ time-domain models on the fitting start time $t_0$, with fixed $M=L=1$, $Q=1.2$, $\eR=-1$, $\ell=2$ and $t_1=35r_h$. The upper panel shows the normalized residual and the lower panel the design-matrix condition number.}
\label{fig:fit-window}
\end{figure}

Table~\ref{tab:td-fits} shows that \texttt{2QNM} and \texttt{EP total} both reach NRMS values near $10^{-10}$. This agreement is expected because the finite-$N$ pair retains a small splitting. The two fits differ strongly in conditioning. The condition numbers are $4.63\times10^{4}$ for \texttt{2QNM} and $1.19\times10^{2}$ for \texttt{EP total}. The latter also uses fewer parameters and is preferred by AIC and BIC for this signal. By contrast, \texttt{EP linear} gives an NRMS of $8.37\times10^{-3}$. The constant Jordan term is therefore required in this example. Figure~\ref{fig:three-fits} shows the fits, and Fig.~\ref{fig:fit-window} shows their dependence on the fitting window.

\begin{table}[htbp]
\centering
\caption{Three fits to the $A=1$ EP two-mode signal at $N=42$ over $12\le t/r_h\le35$, with fixed $M=L=1$, $Q=1.2$, $\eR=-1$ and $\ell=2$. NRMS is the normalized root-mean-square residual and $\kappa_2(X)$ is the 2-norm condition number of the design matrix. The numbers of free real parameters are 8 for \texttt{2QNM}, 4 for \texttt{EP linear}, and 6 for \texttt{EP total}.}
\label{tab:td-fits}
\begin{tabular}{lrrrr}
\toprule
Model & NRMS & $\kappa_2(X)$ & AIC & BIC\\
\midrule
2QNM & $1.11\times10^{-10}$ & $4.63\times10^4$ & $-23906.4$ & $-23873.3$\\
EP linear & $8.37\times10^{-3}$ & $1.19$ & $-7193.9$ & $-7177.4$\\
EP total & $1.08\times10^{-10}$ & $1.19\times10^2$ & $-23942.3$ & $-23917.5$\\
\bottomrule
\end{tabular}
\end{table}

\begin{figure}[htbp]
\centering
\includegraphics[width=0.98\textwidth]{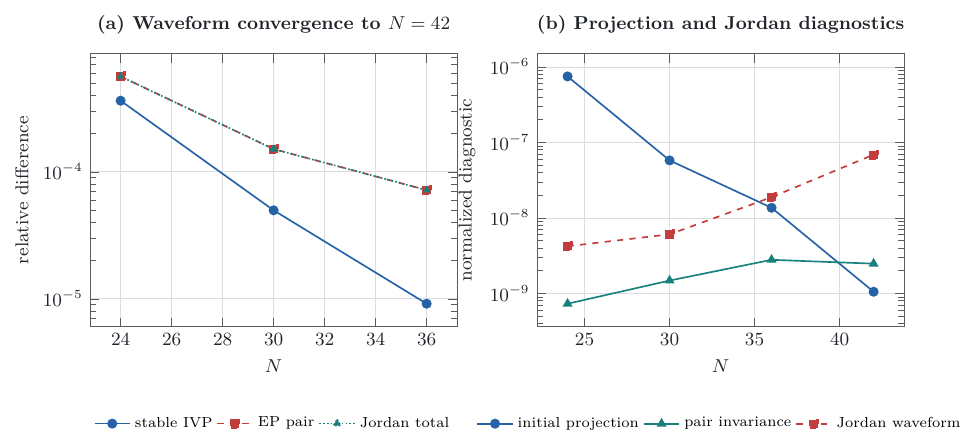}
\caption{Time-domain resolution comparison for the $A=1$ branch at $N=24,30,36,42$, with fixed $M=L=1$, $Q=1.2$, $\eR=-1$ and $\ell=2$. Panel (a) shows the relative changes of the stable IVP, EP pair and Jordan total from $N=42$. Panel (b) shows the change in the initial projection, the pair-invariance diagnostic and the late-time Jordan-waveform reconstruction error.}
\label{fig:td-convergence}
\end{figure}

Figure~\ref{fig:td-convergence} compares independently relocated candidates at $N=24,30,36,42$. Relative to $N=42$, the stable waveform changes by $3.62\times10^{-4}$ at $N=24$ and $9.13\times10^{-6}$ at $N=36$. At higher resolution, double precision no longer resolves the tiny splitting reliably. The reconstruction error is therefore nonmonotonic. Because each resolution uses its own relocated candidate, this comparison is a semidiscrete consistency check, not a fixed-coordinate continuum test.

\clearpage
\section{\texorpdfstring{$A=2$ branch: finite-dimensional frequency-domain results}{A=2 branch: finite-dimensional frequency-domain results}}
\label{sec:a2-frequency}

We then consider the logarithmic $A=2$ branch. Its string-fluid correction decays more slowly than the Newtonian term and changes the potential far from its peak. We take
\begin{equation}
 M=L=\lambda_{\mathrm R}=1,\qquad Q=1,\qquad \eR=+1,\qquad \ell=2,
 \label{eq:a2-parameters}
\end{equation}
For these parameters,
\begin{equation}
 f_{A=2}(r)=1-\frac{2M}{r}+\frac{Q^2}{r^2}
 -\eR\frac{L}{r}\log(\lambda_{\mathrm R}r).
 \label{eq:a2-lapse}
\end{equation}
Here $r_*$ is obtained by numerical integration, with $r_*/r_h=0$ at $r=2r_h$. The value of $a$ therefore uses a gauge convention that differs from $A=1$. The endpoint nonanalyticity and the numerical tortoise coordinate control the convergence. Appendix~\ref{app:a2-reconstruction} gives the coefficient construction and matrix assembly.

\subsection{Two-dimensional localization and the interacting pair}

\begin{figure}[p]
\centering
\begin{minipage}[t]{0.78\textwidth}
\centering
\includegraphics[width=\linewidth]{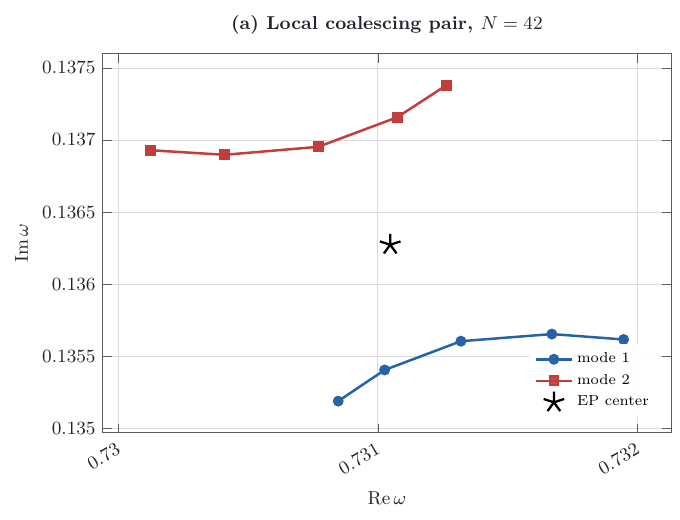}
\end{minipage}\hfill
\par\vspace{0.7em}
\begin{minipage}[t]{0.78\textwidth}
\centering
\includegraphics[width=\linewidth]{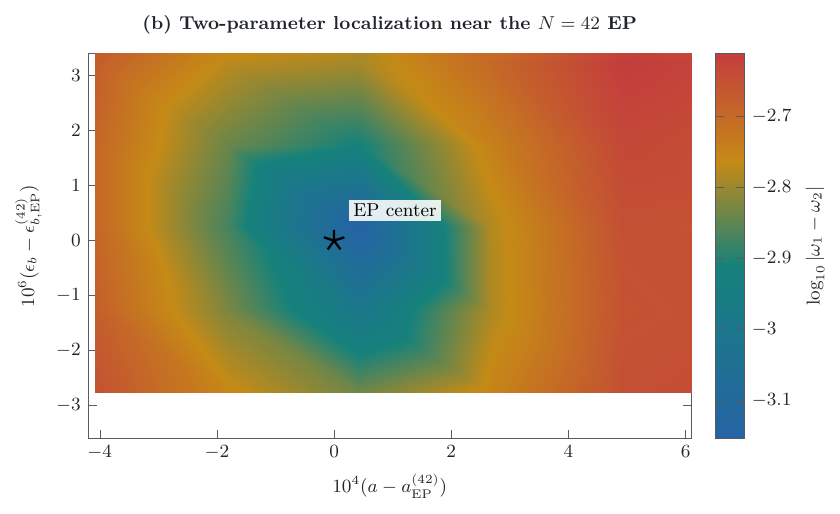}
\end{minipage}
\caption{Local frequency structure of the $A=2$ branch at $N=42$, with fixed $M=L=\lambda_{\mathrm R}=Q=1$, $\eR=+1$ and $\ell=2$. (a) The local coalescing pair of continuously tracked complex-frequency branches; the star marks the center frequency in Eq.~\eqref{eq:a2-point}. (b) Two-parameter localization near the EP, with color denoting $\log_{10}|\omega_1-\omega_2|$; the star marks the finite-dimensional EP center relocated with 128-bit arithmetic, and the heat map is interpolated from a $5\times5$ search grid.}
\label{fig:a2-localization}
\end{figure}

We again use $(a,\eb)$ as the control parameters and define
\begin{equation}
 D_{A=2}(a,\eb)=\left[\omega_1(a,\eb)-\omega_2(a,\eb)\right]^2,
 \label{eq:a2-discriminant}
\end{equation}
We solve $\operatorname{Re}D_{A=2}=\operatorname{Im}D_{A=2}=0$ at $N=42$. Figure~\ref{fig:a2-localization}(b) shows the local gap map. The gap decreases towards an isolated minimum, and the marked point is relocated independently at 128-bit precision.

At $N=42$, arbitrary-precision relocation gives
\begin{subequations}
\label{eq:a2-point}
\begin{align}
 a_{42}&=10.0978081242540521,\\
 \epsilon_{\mathrm b,42}&=0.00943525903135856644,\\
 \omega_{42}&=0.731048009592213085
 +0.136272829358409495\,\ii,
\end{align}
\end{subequations}
The residual gap is $5.33\times10^{-17}$. These digits specify the finite-dimensional $N=42$ problem, not a continuum value. Only the first four or five digits remain stable across resolutions. Figure~\ref{fig:a2-localization}(a) shows the two branches on a slice near $\epsilon_{\mathrm b,42}$.

\subsection{Defectiveness, the Jordan chain, and mode exchange}

We apply the same diagnostics as for $A=1$. With $\bm K(s)=\bm L-s\bm B$ evaluated at $s_{\rm EP}$, the Jordan chain satisfies
\begin{equation}
 \bm K_{\rm EP}\bm u_0=0,\qquad \bm K_{\rm EP}\bm u_1=\bm B\bm u_0,
 \label{eq:a2-jordan}
\end{equation}
At $N=42$ and 128 bits, $1-C_R=2.74\times10^{-32}$, $\widehat\sigma_{\min}=2.50\times10^{-42}$, $r_0=1.62\times10^{-39}$, $r_1=3.40\times10^{-40}$, and $\chi_0=7.17\times10^{-41}$. The largest eigen-equation residual is $6.67\times10^{-37}$. These values show defectiveness of the $N=42$ pencil.

\begin{figure}[htbp]
\centering
\includegraphics[width=0.94\textwidth]{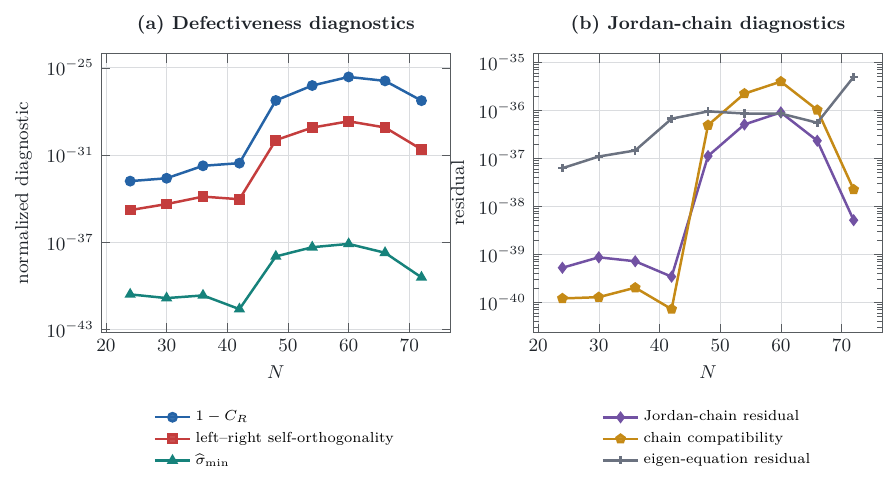}
\caption{Defectiveness and Jordan-chain diagnostics of the $A=2$ branch for $N=24$--$72$, with fixed $M=L=\lambda_{\mathrm R}=Q=1$, $\eR=+1$ and $\ell=2$. Panel (a) shows $1-C_R$, the left--right self-orthogonality measure $R$, and the normalized smallest singular value $\widehat\sigma_{\min}$. Panel (b) shows the Jordan-chain residual $r_1$, the chain compatibility $\chi_0$, and the eigen-equation residual. Each point uses the EP coordinates relocated independently at that resolution.}
\label{fig:a2-jordan}
\end{figure}

Figure~\ref{fig:a2-jordan} collects the diagnostics from $N=24$ to $N=72$. Each resolution shows a defective pencil at its own relocated coordinates. The residuals fluctuate with the relocation tolerance and matrix conditioning. Their smallness establishes the finite-dimensional structure only. Continuum convergence is tested next.

We also test the topology at fixed $N$. In the double-precision $N=42$ calculation, we track the two modes around
\begin{equation}
 a=a_c+0.002\cos\theta,\qquad
 \eb=\ebc+5\times10^{-6}\sin\theta,
 \qquad 0\leq\theta\leq2\pi,
 \label{eq:a2-loop}
\end{equation}
The loop is centered on the double-precision candidate
\begin{equation}
 (a_c,\ebc)=(10.0978080737420,0.00943525976562511),
 \label{eq:a2-loop-center}
\end{equation}
which lies close to the arbitrary-precision coordinates in Eq.~\eqref{eq:a2-point}. After one circuit,
\begin{equation}
 d_{\mathrm{direct}}=8.99\times10^{-3},\qquad
 d_{\mathrm{swapped}}=0,
 \label{eq:a2-loop-result}
\end{equation}
The modes exchange after one circuit and recover their original labels after two, as required for a square-root branch point. Since the loop and tracking use fixed $N$, this result concerns the discrete spectrum only and does not constrain the continuum limit.

\begin{figure}[htbp]
\centering
\includegraphics[width=0.94\textwidth]{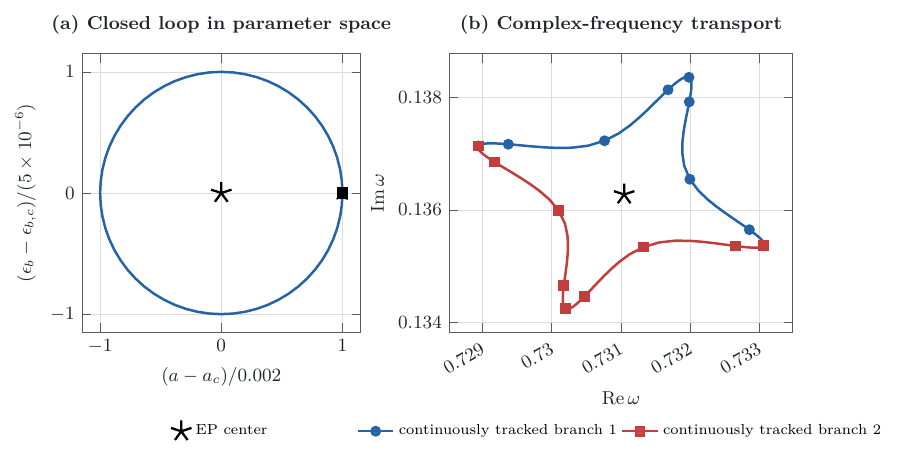}
\caption{Closed-loop transport of the $A=2$ branch at $N=42$, with fixed $M=L=\lambda_{\mathrm R}=Q=1$, $\eR=+1$ and $\ell=2$. (a) The loop in the $(a,\eb)$ plane, with coordinates normalized by the semiaxes $0.002$ and $5\times10^{-6}$. (b) The continuously tracked complex-frequency trajectories. Arrows indicate the transport direction, and the star marks the finite-dimensional EP center. One circuit exchanges the two modes.}
\label{fig:a2-loop}
\end{figure}

\subsection{Resolution dependence}
\label{sec:a2-resolution}

\begin{figure}[htbp]
\centering
\begin{tabular}{@{}cc@{}}
\begin{minipage}[t]{0.46\textwidth}
\centering
\includegraphics[width=\linewidth]{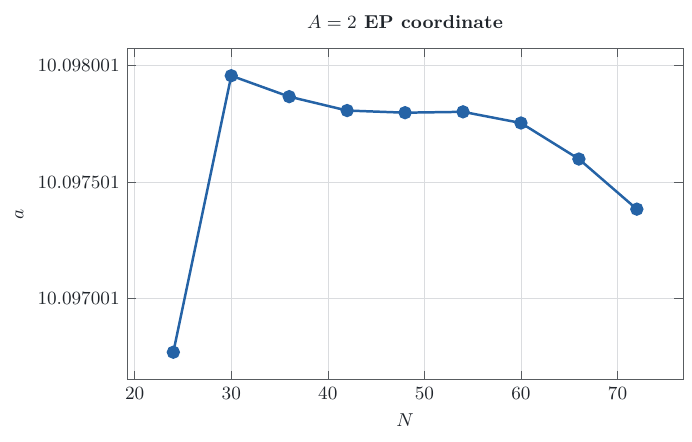}
\end{minipage} &
\begin{minipage}[t]{0.46\textwidth}
\centering
\includegraphics[width=\linewidth]{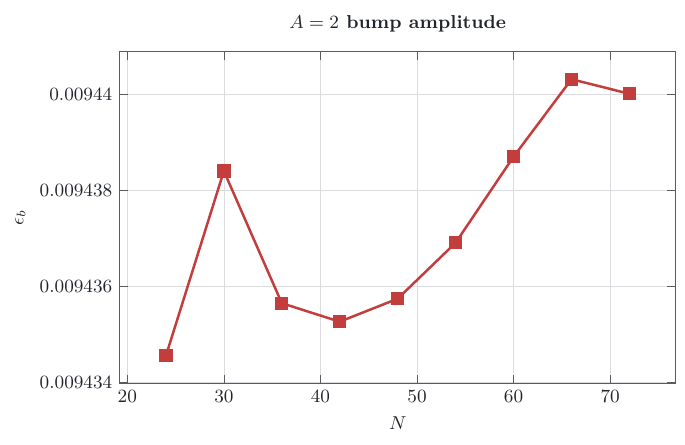}
\end{minipage}\\[0.7em]
\begin{minipage}[t]{0.46\textwidth}
\centering
\includegraphics[width=\linewidth]{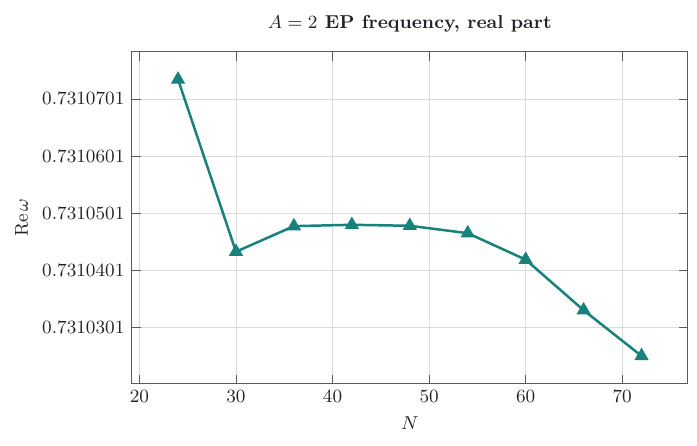}
\end{minipage} &
\begin{minipage}[t]{0.46\textwidth}
\centering
\includegraphics[width=\linewidth]{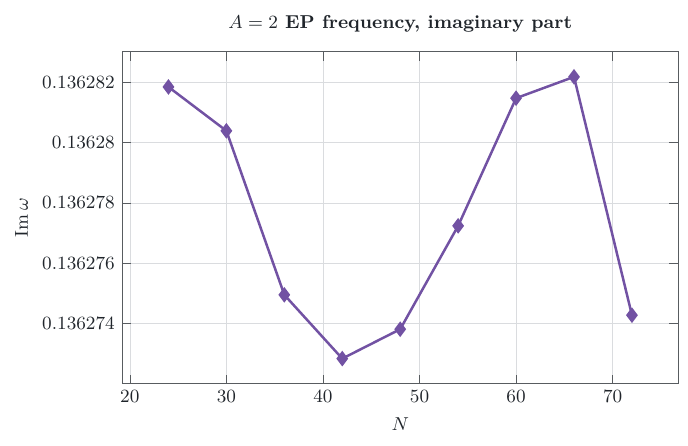}
\end{minipage}
\end{tabular}
\caption{Resolution dependence of independently relocated coordinates for the $A=2$ branch, with fixed $M=L=\lambda_{\mathrm R}=Q=1$, $\eR=+1$ and $\ell=2$. (a) Barrier position $a_N$. (b) Bump amplitude $\epsilon_{b,N}$. (c) Real part of the center frequency, $\operatorname{Re}\omega_N$. (d) Imaginary part, $\operatorname{Im}\omega_N$. Each point is relocated independently at its resolution, and solid lines connect the points in resolution order. The absence of stabilization prevents a continuum interpretation.}
\label{fig:a2-resolution}
\end{figure}

Every resolution gives a defective finite-dimensional EP in the frequency-domain diagnostics. The mode exchange is tested at $N=42$. A continuum interpretation additionally requires the relocated coordinates to stabilize with $N$. Figure~\ref{fig:a2-resolution}(a)--(d) shows the independently relocated $a$, $\eb$, $\operatorname{Re}\omega$, and $\operatorname{Im}\omega$ for $N=24$--$72$. From $N=60$ to $66$,
\begin{align}
 |\Delta a|&=1.55\times10^{-4},&
 |\Delta\eb|&=1.61\times10^{-6},\nonumber\\
 |\Delta\operatorname{Re}\omega|&=8.85\times10^{-6},&
 |\Delta\operatorname{Im}\omega|&=7.07\times10^{-7}.
\end{align}
From $N=66$ to $72$,
\begin{align}
 |\Delta a|&=2.15\times10^{-4},&
 |\Delta\eb|&=3.03\times10^{-7},\nonumber\\
 |\Delta\operatorname{Re}\omega|&=8.00\times10^{-6},&
 |\Delta\operatorname{Im}\omega|&=7.90\times10^{-6}.
\end{align}
For comparison, the $A=1$ sequence reduced successive changes in $a_N$ by about one order of magnitude per step and reached $8.89\times10^{-12}$. In the present branch, $|\Delta a|$ and $|\Delta\operatorname{Re}\omega|$ retain the same order of magnitude, while $|\Delta a|$ increases in the second step. The sequence has not reached a stable regime, so we do not report continuum coordinates.

Two features of the logarithmic branch are consistent with this behavior. First, the endpoint loses analyticity. With $r=r_h/\sigma$,
\begin{equation}
 \sigma^2 f(r_h/\sigma)=\sigma^2-\frac{2M}{r_h}\sigma^3
 +\frac{Q^2}{r_h^2}\sigma^4
 -\eR\frac{L}{r_h}\sigma^3
 \left[\log(\lambda_{\mathrm R}r_h)-\log\sigma\right],
 \label{eq:a2-log-endpoint}
\end{equation}
The $\sigma^3\log\sigma$ term is nonanalytic at future null infinity, $\sigma=0$. Chebyshev approximation is exponentially convergent for analytic functions, but an endpoint branch point reduces the rate to algebraic, as seen in Fig.~\ref{fig:a2-resolution}(a)--(d). By contrast, the $A=1$ coefficients in Eqs.~\eqref{eq:p}--\eqref{eq:qscalar} are polynomial in $\sigma$.

Second, $r_*$ is integrated numerically at the collocation points and inverted at the interface. Equation~\eqref{eq:sqrt} amplifies a coefficient perturbation $\delta$ into a frequency shift of order $O(\sqrt\delta)$. Higher precision removes the rounding floor at fixed $N$, but not the truncation error controlling the resolution trend. The present data do not separate the endpoint and tortoise contributions quantitatively, so we attribute the observed behavior to both.

The two-parameter search, eigenvector coalescence, Jordan chain, and closed-loop exchange identify a second-order defective degeneracy of the discrete problem. The relocated values cluster near $(a,\eb)\simeq(10.0978,0.009437)$ and $\omega\simeq0.73104+0.13628\,\ii$, with Eq.~\eqref{eq:a2-point} as the representative $N=42$ result. Because the $N=24$--$72$ sequence has not stabilized, these values are truncation-dependent and do not define a converged continuum EP. A continuum EP remains plausible, but testing it requires a discretization adapted to the $\sigma^3\log\sigma$ endpoint. We have not performed an equivalent two-parameter search for the $A=2$, $\eR=-1$ branch.

\section{\texorpdfstring{$A=2$ branch: finite-dimensional time-domain response}{A=2 branch: finite-dimensional time-domain response}}
\label{sec:a2-time}

We next test the Jordan response of the finite-dimensional $A=2$ degeneracy. We use $\eR=+1$, $Q=1$, and $\ell=2$. Each resolution has its own relocated $(a_N,\epsilon_{\mathrm b,N},\omega_N)$. We evolve $N=24,30,36,42$ with $\Delta t=0.05$ and the matrix exponential defined in Sec.~\ref{sec:a1-time}. Because the coordinates do not converge, this section concerns the corresponding discrete matrices only.

\subsection{Constraint elimination and a stable initial-value problem}

As in the $A=1$ branch, the interface conditions make $\bm B$ singular. We eliminate them with Eq.~\eqref{eq:embedding} and $\bm C\bm Z=0$, giving
\begin{equation}
 \frac{\dd}{\dd t}\begin{pmatrix}\bm q\\\bm p\end{pmatrix}
 =\bm A_{\rm red}\begin{pmatrix}\bm q\\\bm p\end{pmatrix},
 \qquad
 \bm A_{\rm red}=\begin{pmatrix}\bm0&\bm I\\\bm A_1&\bm A_2\end{pmatrix}.
 \label{eq:a2-reduced-ivp}
\end{equation}
The initial data remain $(\Phi,\partial_t\Phi)|_{t=0}=(1,0)$, and the waveform is read at $\scri^+$. At $N=42$, the constraint residual is $2.98\times10^{-20}$ and the normalized residual in the original equations is $1.93\times10^{-17}$.

\begin{figure}[htbp]
\centering
\includegraphics[width=0.98\textwidth]{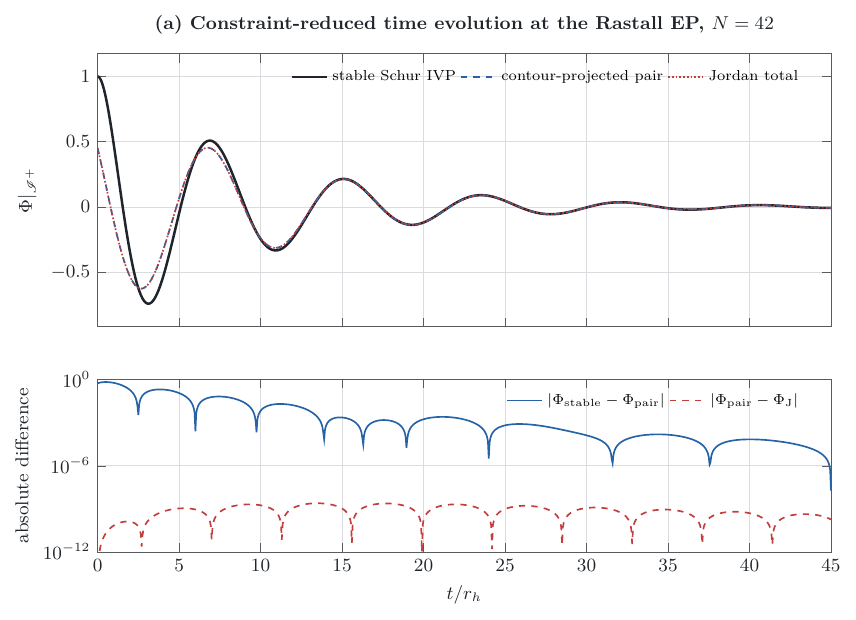}
\caption{Time-domain waveform of the $A=2$ branch at $N=42$, with fixed $M=L=\lambda_{\mathrm R}=Q=1$, $\eR=+1$, $\ell=2$ and the finite-dimensional EP coordinates in Eq.~\eqref{eq:a2-point}. The black solid curve is the initial-value evolution in the stable Schur subspace, the blue dashed curve is the EP two-dimensional contribution, and the red dotted curve is the Jordan reconstruction; the lower panel shows the absolute differences from the EP-pair propagation.}
\label{fig:a2-td-waveform}
\end{figure}

The $A=2$ generator also contains spurious growing directions. At $N=42$, 32 eigenvalues have $\operatorname{Re}s>0$, with a maximum of $4.22\times10^3$. We retain the damped Schur subspace. The projection changes the initial state by $1.15\times10^{-7}$ and has an invariance residual of $1.18\times10^{-12}$. Figure~\ref{fig:a2-td-waveform} shows the resulting waveform and its EP-pair component.

\subsection{Riesz projection, Jordan response, and fits}

At $N=42$, with $s=\ii\omega$, we choose a contour that encloses only the interacting pair and evaluate the Riesz projector in Eq.~\eqref{eq:riesz-projector}. The invariant-subspace residual is $4.91\times10^{-9}$. The ratio of the third to second singular value is $4.19\times10^{-12}$, indicating a two-dimensional EP subspace. With $s_c$ the mean of the enclosed eigenvalues, Eq.~\eqref{eq:jordan-cd} gives
\begin{equation}
 C=0.2250256+0.3111708\,\ii,\qquad
 D=-0.0056158+0.0339254\,\ii,
 \label{eq:a2-jordan-cd}
\end{equation}
The corresponding real response is
\begin{equation}
 \Phi_{\rm J}(t)=2\operatorname{Re}\left[(C+Dt)e^{s_ct}\right].
 \label{eq:a2-jordan-waveform}
\end{equation}
The term $D\,te^{s_ct}$ is the predicted $te^{+\ii\omega_{\rm EP}t}$ contribution. The compressed generator is nilpotent to $1.76\times10^{-10}$. Over $12\le t/r_h\le35$, Eq.~\eqref{eq:a2-jordan-waveform} agrees with the two-dimensional matrix exponential to a relative $1.55\times10^{-8}$. This linear-in-time response is obtained directly from the discrete matrix, without fitting.

\begin{figure}[htbp]
\centering
\includegraphics[width=0.96\textwidth]{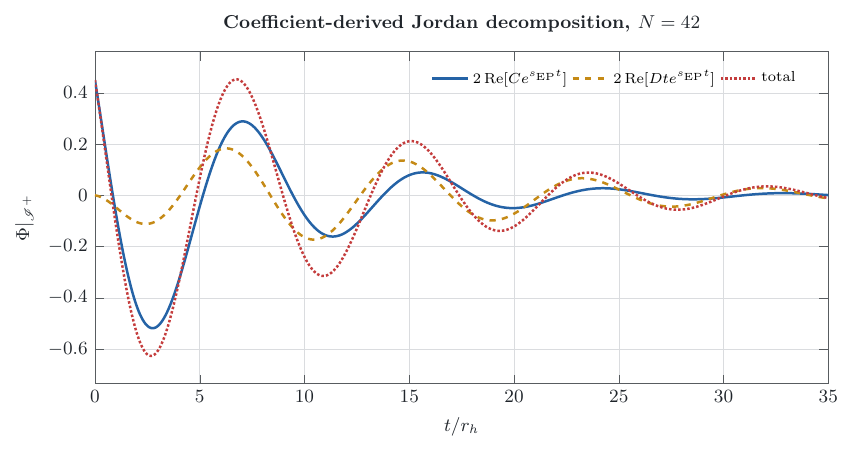}
\caption{Jordan response of the $A=2$ branch at $N=42$, with fixed $M=L=\lambda_{\mathrm R}=Q=1$, $\eR=+1$ and $\ell=2$. The curves show the constant term $2\operatorname{Re}(Ce^{s_ct})$, the linear term $2\operatorname{Re}(Dte^{s_ct})$, and their sum, which contains the characteristic linear-in-time contribution.}
\label{fig:a2-jordan-components}
\end{figure}

We fit the same signal over $12\le t/r_h\le35$ with the three models defined in Sec.~\ref{sec:a1-time}; the results appear in Table~\ref{tab:a2-td-fits}. The \texttt{2QNM} and \texttt{EP total} models both reach NRMS values of order $10^{-10}$. Their condition numbers are $3.47\times10^{4}$ and $9.13\times10^{1}$, a factor of about $3.8\times10^{2}$. The \texttt{EP linear} model remains near $10^{-2}$, so both Jordan terms are needed.

Here the information criteria favor \texttt{2QNM}, unlike the $A=1$ fit. Because the relocated $A=2$ coordinates are not converged, the finite-$N$ pair retains a truncation-dependent splitting that an independent-frequency model can exploit. These criteria therefore do not distinguish continuum representations. The conditioning gap is the more reliable observation for this fixed-$N$ signal.

\begin{figure}[htbp]
\centering
\includegraphics[width=0.98\textwidth]{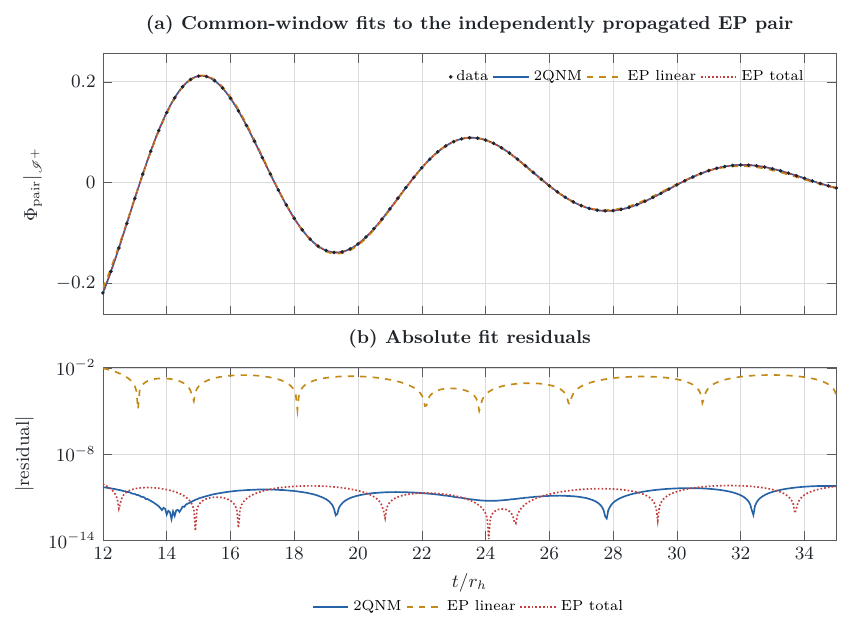}
\caption{Fits of the same $A=2$ EP two-mode signal at $N=42$ with \texttt{2QNM}, \texttt{EP linear} and \texttt{EP total}, for fixed $M=L=\lambda_{\mathrm R}=Q=1$, $\eR=+1$, $\ell=2$ and $12\le t/r_h\le35$. The upper panel shows the waveform and the lower panel the absolute residual.}
\label{fig:a2-three-fits}
\end{figure}

\begin{figure}[htbp]
\centering
\includegraphics[width=0.98\textwidth]{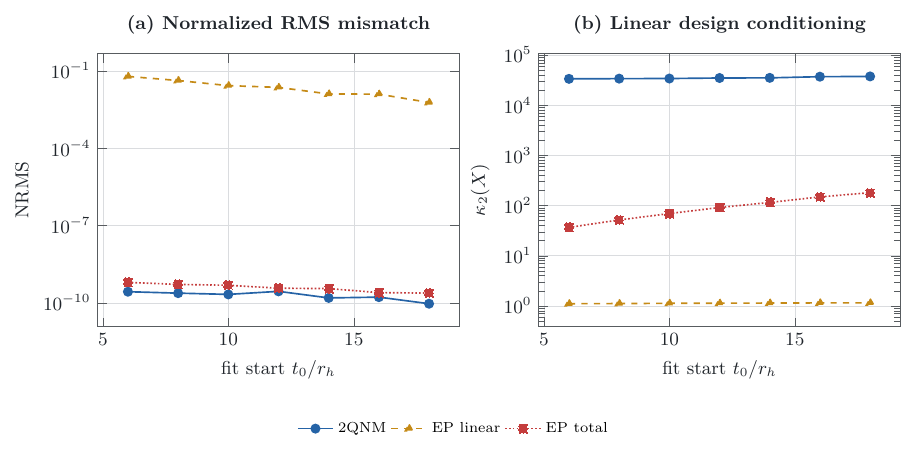}
\caption{Dependence of the three $A=2$ time-domain models on the fitting start time $t_0$, with fixed $M=L=\lambda_{\mathrm R}=Q=1$, $\eR=+1$, $\ell=2$ and $t_1=35r_h$. The upper panel shows the normalized residual and the lower panel the design-matrix condition number.}
\label{fig:a2-fit-window}
\end{figure}

Figure~\ref{fig:a2-three-fits} shows the fits and residuals, while Fig.~\ref{fig:a2-fit-window} shows the dependence on the fitting start time. The \texttt{EP linear} residual remains larger in every window. The other two residuals stay low, and their condition-number separation persists.
\begin{table}[htbp]
\centering
\caption{Three fits to the $A=2$ EP two-mode signal at $N=42$ over $12\le t/r_h\le35$, with fixed $M=L=\lambda_{\mathrm R}=Q=1$, $\eR=+1$ and $\ell=2$. NRMS is the normalized root-mean-square residual and $\kappa_2(X)$ is the 2-norm condition number of the design matrix. The numbers of free real parameters are 8 for \texttt{2QNM}, 4 for \texttt{EP linear}, and 6 for \texttt{EP total}.}
\label{tab:a2-td-fits}
\begin{tabular}{lrrrr}
\toprule
Model & NRMS & $\kappa_2(X)$ & AIC & BIC\\
\midrule
2QNM & $2.81\times10^{-10}$ & $3.47\times10^4$ & $-22502.3$ & $-22469.3$\\
EP linear & $2.29\times10^{-2}$ & $1.13$ & $-5714.3$ & $-5697.8$\\
EP total & $3.71\times10^{-10}$ & $9.13\times10^1$ & $-22248.7$ & $-22223.9$\\
\bottomrule
\end{tabular}
\end{table}

\subsection{Resolution test of the time-domain result}

\begin{figure}[htbp]
\centering
\includegraphics[width=0.98\textwidth]{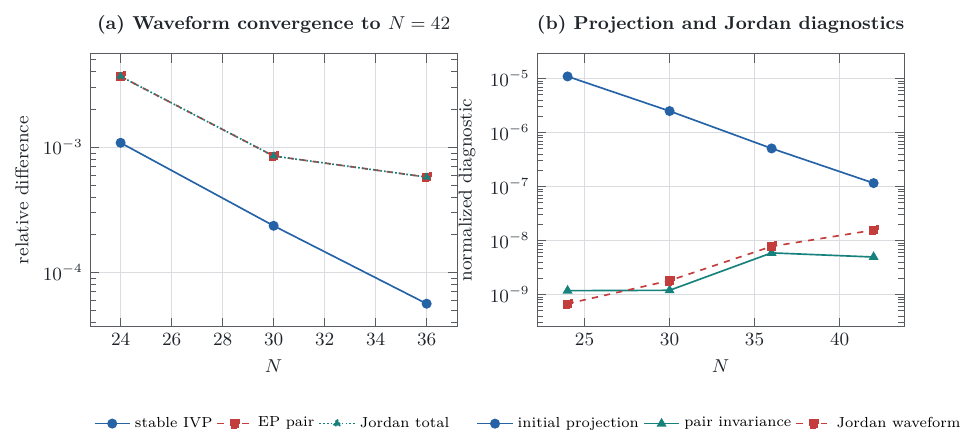}
\caption{Time-domain resolution comparison for the $A=2$ branch at $N=24,30,36,42$, with fixed $M=L=\lambda_{\mathrm R}=Q=1$, $\eR=+1$ and $\ell=2$. Each resolution uses its own relocated $(a_N,\epsilon_{\mathrm b,N},\omega_N)$, so this is not a fixed-parameter continuum waveform-convergence test. Panel (a) shows the relative changes of the stable IVP, EP pair and Jordan total from $N=42$. Panel (b) shows the change in the initial projection, the pair-invariance diagnostic and the late-time Jordan-waveform reconstruction error.}
\label{fig:a2-td-convergence}
\end{figure}

Figure~\ref{fig:a2-td-convergence} collects the $N=24,30,36,42$ calculations. Relative to $N=42$, the stable initial-value waveform changes by $1.08\times10^{-3}$ at $N=24$ and $5.61\times10^{-5}$ at $N=36$. The EP two-mode component changes by $3.66\times10^{-3}$ and $5.75\times10^{-4}$, respectively. At $N=42$, the late-time Jordan reconstruction differs by $1.55\times10^{-8}$.

The changes are not monotonic, reflecting the double-precision Riesz integral, square-root sensitivity near the degeneracy, and matrix-exponential evaluation. Higher precision would be required to reduce each contribution. The matrix in Eq.~\eqref{eq:a2-point} nevertheless shows a Jordan-consistent response and the $te^{+\ii\omega_{\rm EP}t}$ term under constrained evolution. Because the coordinates do not stabilize, this waveform is not a continuum fixed-parameter test.

\section{Conclusion}
\label{sec:conclusion}

In this paper, we have studied non-Hermitian degeneracies in the scalar QNM spectrum of Rastall string-fluid black holes by introducing a localized bump whose position and amplitude provide the two real controls required for an isolated second-order EP. For the $A=1$, $\eR=-1$ branch, arbitrary-precision relocation from $N=24$ to $N=60$ gives the coordinates in Eqs.~\eqref{eq:aep}--\eqref{eq:omegaep}, while the successive changes decrease by about one order of magnitude per refinement step. Eigenvector coalescence, vanishing left--right self-orthogonality, a length-two Jordan chain and one-loop mode exchange all support defectiveness, and the local splitting follows the $1/2$ exponent in Eq.~\eqref{eq:sqrt} on both sides of the candidate.

The constrained initial-value problem isolates the EP subspace with a Riesz projector and produces the $te^{+\ii\omega_{\rm EP}t}$ term. The Jordan reconstruction agrees with two-dimensional propagation to $6.87\times10^{-8}$ over $12\le t/r_h\le35$, and this term comes from the nilpotent part of the compressed generator rather than from a fitting template. The EP-pair contribution differs from the full stable waveform by $2.97\times10^{-2}$, so the agreement is an internal semidiscrete check rather than an independent continuum validation. For the logarithmic $A=2$, $\eR=+1$ branch, defectiveness appears in every truncation from $N=24$ to $N=72$, mode exchange is demonstrated at $N=42$, and the representative values in Eq.~\eqref{eq:a2-point} reconstruct the Jordan signal to $1.55\times10^{-8}$. The relocated coordinates stop converging beyond $N=60$, likely because of the $\sigma^3\log\sigma$ endpoint nonanalyticity, the numerical construction of $r_*$ and the square-root sensitivity near the degeneracy. We therefore report the $A=2$ case as a finite-dimensional result: a continuum EP remains plausible, but it is not established.

Two observations are relevant to ringdown modeling. The condition-number gap between the \texttt{2QNM} and \texttt{EP total} parameterizations is about $390$ in both branch-specific signals, although this factor is limited to the present fits. Nearly degenerate exponentials can generally be ill-conditioned. The failure of \texttt{EP linear} further shows that a complete EP response needs the constant Jordan term together with the linear term. These results agree with recent near-EP ringdown analyses \cite{PanossoMacedo:2025xnf,Kubota:2025hjk}, but they come from a scalar test field, a Rastall string-fluid background and a phenomenological bump, so they do not establish a Rastall-specific EP mechanism. The calculation is limited to one background parameter set and one low-lying scalar mode pair, and the bump in Eq.~\eqref{eq:bump} is not a solution of the Rastall field equations; its coordinates are therefore not predictions for a specific astrophysical environment.

The calculation establishes that the discretized spectrum admits second-order defective degeneracies and that two-parameter searches can locate them to high precision. Future work can examine additional spectral and perturbative sectors, such as continuation in $Q$ and $\ell$, higher mode pairs, electromagnetic and gravitational perturbations, and exceptional-line organization \cite{Cao:2025afs,Cavalcante:2026exl}, etc. For the $A=2$ branch, a discretization adapted to the $\sigma^3\log\sigma$ endpoint and an analytic treatment of $r_*$ are needed before a continuum statement can be made.

\appendix
\section{Height function and endpoint regularity}

A minimal-gauge height function compatible with Eq.~\eqref{eq:gamma} is
\begin{align}
 H(\sigma)={}&-\frac1\sigma+(1+\kappa)\log\sigma
 +\frac{\log(1-\sigma)}{1-\kappa}\nonumber\\
 &+\frac{\kappa^2}{1-\kappa}\log(1-\kappa\sigma),
\end{align}
and it satisfies $\gamma=pH'$. For $\kappa\in(-1,1)$, Eqs.~\eqref{eq:tortoise} and the endpoint asymptotics give super-exponential decay of the bump as $\sigma\to0$ and $\sech^2(x-a)/p\sim(1-\sigma)^{2/(1-\kappa)-1}\to0$ as $\sigma\to1$. Neither physical boundary therefore requires an artificial bump cutoff.

\section{\texorpdfstring{$A=2$ operator and matrix reconstruction}{A=2 operator and matrix reconstruction}}
\label{app:a2-reconstruction}

This appendix specifies the finite-dimensional problem used for the logarithmic branch. We retain $\sigma=r_h/r$ and define
\begin{align}
 p_{2}(\sigma)&=\sigma^2 f_{A=2}(r_h/\sigma),&
 \gamma_{2}(\sigma)&=1-2\sigma^2,\\
 w_{2}(\sigma)&=\frac{1-\gamma_{2}(\sigma)^2}{p_{2}(\sigma)},&
 q_{s,2}(\sigma)&=\ell(\ell+1)
 +\frac{r_h f'_{A=2}(r_h/\sigma)}{\sigma}.
\label{eq:a2-coefficients}
\end{align}
The endpoint limits used in the implementation are $w_2(0)=4$ and $w_2(1)=8/[r_h f'_{A=2}(r_h)]$. The bump enters the total coefficient as
\begin{equation}
 q_{2}(\sigma)=q_{s,2}(\sigma)+
 \eb\,\frac{\sech^2[x(\sigma)-a]}{p_{2}(\sigma)}.
\label{eq:a2-qtotal}
\end{equation}

The additive constant of the tortoise coordinate is fixed by $x=0$ at $r=2r_h$. For numerical integration we use
\begin{equation}
 u=\log\frac{1-\sigma}{\sigma},\qquad
 r=r_h(1+e^u),\qquad
 \frac{\dd x}{\dd u}=
 \frac{e^u}{f_{A=2}\!\left[r_h(1+e^u)\right]},
\label{eq:a2-tortoise-integration}
\end{equation}
and integrate from $u=0$. The Float64 implementation uses adaptive Simpson quadrature with $\mathrm{rtol}=2\times10^{-11}$, $\mathrm{atol}=2\times10^{-12}$ and maximum recursion depth 22. The interface $x(\sigma_b)=a$ is obtained by bisection on $\sigma\in[10^{-12},0.5]$ for 80 iterations. The arbitrary-precision implementation uses 48-point Gauss--Legendre quadrature on $u$ panels no wider than 1.5 and solves for the interface in $u$ by Newton iteration for 30 iterations with tolerance $\sqrt{\epsilon(T)}$ at the working type.

For each domain, the first-order operators are
\begin{equation}
 \mathcal L_{1,2}=p_2\partial_\sigma^2+p_2'\partial_\sigma-q_2,
 \qquad
 \mathcal L_{2,2}=2\gamma_2\partial_\sigma+\gamma_2',
\label{eq:a2-first-order}
\end{equation}
and the bulk matrix is assembled in the state ordering $(\bm\Phi_0,\bm\Psi_0,\bm\Phi_1,\bm\Psi_1)$:
\begin{equation}
 \bm L_{\rm bulk}=
 \begin{pmatrix}
 0&I&0&0\\
 \mathcal L_{1,2}/w_2&\mathcal L_{2,2}/w_2&0&0\\
 0&0&0&I\\
 0&0&\mathcal L_{1,2}/w_2&\mathcal L_{2,2}/w_2
 \end{pmatrix},
 \qquad \bm B_{\rm bulk}=I.
\label{eq:a2-matrix-block}
\end{equation}
The two interface rows replace the corresponding bulk equations and set the same rows of $\bm B$ to zero:
\begin{equation}
 \Phi_0(\sigma_b)-\Phi_1(\sigma_b)=0,\qquad
 \partial_\sigma\Phi_0(\sigma_b)-\partial_\sigma\Phi_1(\sigma_b)=0.
\label{eq:a2-interface}
\end{equation}
The hyperboloidal endpoint treatment is inherited from the regularized first-order formulation. No additional bump cutoff or external boundary condition is introduced at either physical endpoint.

For an $N=42$ reconstruction, the published finite-dimensional input is
\begin{center}
\begin{tabular}{ll}
\toprule
Background & $M=L=\lambda_{\mathrm R}=Q=1,\ \eR=+1,\ \ell=2$\\
Arithmetic & 128-bit arbitrary precision\\
Coordinate gauge & $x=0$ at $r=2r_h$\\
EP coordinates & $a=10.0978081242540521,\quad
 \eb=0.00943525903135856644$\\
Center frequency & $\omega=0.731048009592213085+
 0.136272829358409495\,\ii$\\
Frequency gap & $5.33\times10^{-17}$\\
Jordan diagnostics & $r_0=1.62\times10^{-39},\
 r_1=3.40\times10^{-40},\
 \chi_0=7.17\times10^{-41}$\\
Loop diagnostic & $d_{\rm direct}=8.99\times10^{-3},\
 d_{\rm swapped}=0$\\
\bottomrule
\end{tabular}
\end{center}
These values define the reported $N=42$ finite pencil and its diagnostics. They do not specify a continuum EP because the independently relocated coordinates remain resolution dependent.

\section{Numerical implementation and data availability}

The frequency-domain search uses two-domain Chebyshev--Lobatto collocation, matching principal and reference resolutions, and frequency continuation between neighboring parameter points. These steps reject migrating pseudospectral points during mode tracking.

Arbitrary-precision verification uses shift-invert full-spectrum solves at the precisions listed in Sec.~\ref{sec:model-method}, SVD-based Jordan-chain construction, and independent relocation at each $N$. The time-domain calculation eliminates the interface constraints, projects onto the damped complex-Schur subspace, and extracts the EP subspace with a 96-point trapezoidal Riesz integral. Appendix~\ref{app:a2-reconstruction} gives the $A=2$ coefficient formulas, quadrature settings, interface inversion, matrix blocks, and $N=42$ input.

The numerical outputs and calculation data are available from the authors upon reasonable request.

 \vspace{1cm}
{\bf Acknowledgments}
 \vspace{0.5cm}

This research is supported by the Natural Science Basic Research Program of Shaanxi Province (Program Nos. 2026JC-YBMS-0083 and 2026JC-YBMS-0101), the National Natural Science Foundation of China (NNSFC) (Grant No.12365009), and the Jiangxi Provincial Natural Science Foundation (No.20262BAC240347).

\bibliographystyle{unsrt}
\bibliography{refs}

\end{document}